\documentclass[%
reprint,
twocolumn,
superscriptaddress,
nofootinbib,
 amsmath,amssymb,
 aps,
]{revtex4-1}
\usepackage{graphicx}
\usepackage{dcolumn}
\usepackage{array}
\usepackage{hyperref}
\usepackage{bm}
\usepackage{xcolor}
\usepackage{enumitem}
\usepackage{aas_macros}  
\usepackage[export]{adjustbox}

\begin{document}

\newcommand{\gf}[1]{{\bf\color{cyan}{[GF: #1]}}}
\newcommand{\xf}[1]{{\bf\color{purple}{[XXK: #1]}}}


\preprint{APS/123-QED}

\title{On the Stochastic Gravitational Wave Background from Binary Black Hole Mergers Dynamically Assembled in Dense Star Clusters}

\author{Xiao-Xiao Kou}
\email{kou00016@umn.edu}
\affiliation{%
University of Minnesota, School of Physics and Astronomy, Minneapolis, MN 55455, USA}%
\author{Giacomo Fragione}%
\affiliation{%
Center for Interdisciplinary Exploration and Research in Astrophysics (CIERA), 1800 Sherman, Evanston, IL 60201, USA
}
\affiliation{Department of Physics \& Astronomy, Northwestern University, Evanston, IL 60208, USA}
\author{Vuk Mandic}
\affiliation{University of Minnesota, School of Physics and Astronomy, Minneapolis, MN 55455, USA}


\begin{abstract}
With about a hundred binary black hole (BBH) mergers detected by LIGO-Virgo-KAGRA, and with several hundreds expected in the current O4 run, GWs are revolutionizing our understanding of the universe. Some BBH sources are too faint to be individually detected, but collectively they may give rise to a stochastic GW background (SGWB). In this paper, we calculate the SGWB associated with BBH mergers dynamically assembled in dense star clusters, using state-of-the-art numerical models. We discuss the role of modeling the evolution of the mass distribution of BBH mergers, which has significant implications for model selection and parameter estimation, and could be used to distinguish between different channels of BBH formation. We demonstrate how the birth properties of star clusters affect the amplitude and frequency spectrum of the SGWB, and show that upcoming observation runs of ground-based GW detectors may be sensitive enough to detect it. Even in the case of a non-detection, we find that GW data can be used to constrain the highly uncertain cluster birth properties, which can complement direct observations of young massive clusters and proto-star clusters in the early universe by JWST.
\end{abstract}

\maketitle

\section{Introduction}
Coalescing binary black holes (BBHs) are among the primary sources of detectable gravitational waves (GWs) for the LIGO~\cite{LIGOScientific:2014pky}, Virgo~\cite{VIRGO:2014yos}, and KAGRA~\cite{KAGRA:2020tym} detectors. The LIGO-Virgo-KAGRA (LVK) Collaborations have reported about a hundred GW event candidates from BBHs when combining data from the LIGO and Virgo observatories during the first three observational runs~\cite{LIGOScientific:2020ibl,LIGOScientific:2021djp,LIGOScientific:2021usb,KAGRA:2021kbb}. These observations have significantly contributed to our understanding of black hole (BH) populations, their formation and merger rates~\cite{Kimball:2020qyd,Zevin:2020gbd,Mapelli:2019bnp,Baibhav:2019gxm,Broekgaarden:2021efa,vanSon:2021zpk}.

There exist various proposed formation channels for BBHs (see, e.g., \cite{Mapelli:2021taw,Mandel:2021smh} for recent reviews). Compact binary mergers can occur through isolated binary evolution within stellar fields~\cite{Belczynski:2001uc,Dominik:2012kk,Belczynski:2017gds} or through dynamical assembly in dense environments, such as globular clusters (GCs), young star clusters (YSCs) or nuclear star clusters (NSCs)~\cite{Rodriguez:2016kxx,Rodriguez:2017pec,DiCarlo:2019pmf,Antonini:2018auk,Mapelli:2021gyv,FragioneKocsisetal2022}. Other channels include hierarchical (triple and quadruple) systems \citep{HoangNaoz2018,FragioneLoebRasio2020}, AGN disks \citep{BartosKocsis2017}, and primordial BHs~\cite{Sasaki:2016jop,Ali-Haimoud:2016mbv,Bird:2016dcv,Clesse:2016vqa}.

One notable event detected during the O3a observation run, GW190521~\cite{LIGOScientific:2020iuh,LIGOScientific:2020ufj}, potentially encompasses one or both component within the pair-instability mass gap, where no BHs are expected to be formed as a result of the death of massive stars~\cite{BelczynskiHeger2016,Spera:2017fyx,Stevenson:2019rcw}. This detection suggests that the system might have formed through successive dynamical encounters in a dense star cluster, indicating that this particular formation channel may contribute to the overall merger rate with a non-negligible fraction~\cite{Fishbach:2020qag,Fragione:2020han,Kimball:2020qyd,Liu:2020gif,FragioneRasio2023}. 

GCs are ubiquitous in our universe and play a key role in astrophysics, from their relationship to galaxy assembly to the exotic stellar populations they generate \citep{2018RSPSA.47470616F,2019A&ARv..27....8G}. However, the precise formation process of GCs, as well as their population properties such as mass, radius, and metallicity distribution, as a function of cosmic time remain poorly constrained \citep{PortegiesZwartMcMillan2010}. GW signals emitted by merging BBHs provide valuable insights not only into the sources themselves but also into the evolutionary characteristics of their host environments \citep{Fishbach:2023xws}, offering an alternative approach to conventional electromagnetic observations~\cite{Vanzella,Lamiya}.

Given the limited sensitivity of current and near-future GW detectors, the redshift distribution of BBHs can only be traced up to $z\leq 2$~\cite{Vitale:2018yhm,KAGRA:2013rdx}. Romero et al.~\cite{Romero-Shaw:2020siz} demonstrated that next-generation GW detectors may expand our understanding of GC formation by measuring the rate of BBH mergers originating therein, extending to redshifts $z \gtrsim 15$. Moreover, recent work by Fishbach et al.~\cite{Fishbach:2023xws} illustrated that GW observations could already provide valuable constraints on GC properties by carefully considering the spin distribution of the GWTC-3 catalog and employing detailed modeling of GC evolution.

While the observed coalescing BBHs are relatively loud and nearby events, many more distant, fainter, and unresolved BBH mergers are expected at higher redshifts, collectively giving rise to the stochastic gravitational-wave background (SGWB). The data collected during the first three observing runs of Advanced LIGO and Advanced Virgo have been utilized to search for the SGWB in the frequency range of approximately 10 to 100\,Hz. However, so far, only upper limits have been established as a result of a non-detection~\cite{KAGRA:2021kbb,LIGOScientific:2019vic,LIGOScientific:2018mvr,LIGOScientific:2016jlg}. The most stringent upper limit, as provided by the LVK collaboration in~\cite{KAGRA:2021kbb}, sets $\Omega_{\mathrm{GW}} \leq 3.4\times 10^{-9}$ at 25 Hz for a power law background with a spectral index of 2/3 ($\Omega_{\mathrm{GW}}$ is the dimensionless energy-density spectrum, see Eq.~\ref{eq:five}). It is anticipated that the SGWB will be detected a few years after the second-generation (2G) detectors have reached their design sensitivities~\cite{KAGRA:2021kbb,KAGRA:2021duu}. 
 
In this paper, our main goal is to characterize the contribution of coalescing BBHs dynamically assembled in dense star clusters to the SGWB, and to determine how the distributions of cluster properties at birth affect the SGWB amplitude and frequency spectrum. Importantly, our study represents a significant advancement compared to earlier studies conducted by \cite{Zhao:2020iew,Perigois:2021ovr,Bavera:2021wmw}. Instead of relying solely on phenomenological astrophysical models or simple prescriptions for the population of dense star clusters (see \cite{Zevin:2020gbd} for details), we incorporate the results of state-of-the-art GC simulations within a self-contained framework, which accounts for different distribution of their masses, radii, formation times, and metallicities. This approach enables us to achieve a more realistic characterization of the evolution of GCs, and capture the dependence between the SGWB and their host environment. We also explore how the variation in the distribution of BBH mass components with respect to redshift can result in different SGWB energy density spectra, which can improve our ability to distinguish between different astrophysical channels for the origin of BBH mergers. Finally, we discuss the importance of considering potential impacts of modeling systematics, such as the BBH merger rate and mass distribution evolution, when conducting isotropic SGWB searches using real data. 

The paper is structured as follows. In Section~\ref{sec:method}, we provide a description of our methodology, which includes modeling BBH mergers along with the evolution of dense star clusters and the characteristics of their SGWB energy spectrum. Our main results are presented in Section~\ref{sec:result}, where we demonstrate how the detection, or a lack of detection, of the SGWB can be leveraged to further constrain the parameter space of GC properties at birth. We also investigate how considering the redshift evolution of the distribution of BBH mass components can enhance the differentiation of GC models and can help reduce biases in parameter estimation, highlighting the importance of accounting for the SGWB modeling systematics. We discuss the implications of our findings and summarize directions for future work in Section~\ref{sec:summary}.

\section{Methods}
\label{sec:method}
\subsection{Globular cluster models}
To assess the population size of BBH mergers originating from GCs, we use data from the \textsc{cluster monte carlo} (\textsc{cmc}) catalog \cite{Kremer:2019iul}. Details of the H\'enon-type Monte Carlo algorithm that is used to study the long-term evolution of the GCs can be found in \cite{Rodriguez:2021qhl}. Here, we summarize some of the most important features of the models.

The \textsc{cmc} catalog includes a diverse set of initial conditions, encompassing variations in several key properties to construct a comprehensive grid of GC models. These properties consist of the following: the overall count of single stars and binaries within the cluster, taking values of $ N = 2\times 10^5$, $4\times 10^5$, $8\times 10^5$, and $1.6\times 10^6$; the initial virial radius of the cluster encompassing values of $r_v = 0.5, 1, 2, 4 ~\mathrm{pc}$; the galactocentric radius of the cluster featuring values of $R_{\mathrm{gc}} = 2, 8, 20 ~\mathrm{kpc}$; and the initial metallicity of the cluster $Z= 0.01, 0.1, 1~Z_{\odot}$. Finally, the catalog includes also four additional clusters with the highest masses, characterized by $N=3.2\times 10^6$, two metallicities ($Z=0.01~Z_{\odot}, 1~Z_{\odot}$), and two radii ($r_v = 1, 2$~$\mathrm{pc}$), while maintaining a fixed galactocentric radius ($R_{\mathrm{gc}} = 20 ~\mathrm{kpc}$). As a result, we have a total number of $148$ different cluster models.

The \textsc{cmc} cluster models are assumed to be initially described by a King profile \cite{King1966} with a concentration parameter $W_0=5$. Individual stellar masses are sampled from a canonical Kroupa initial mass function in the range $[0.08-150]~M_{\odot}$ \cite{Kroupa2001}, with an initial binary fraction of $5\%$. Secondary masses in binaries are sampled from a uniform distribution in mass ratio \cite{DuquennoyMayor1991}, while binary orbital periods are sampled from a log-uniform distribution from near contact to the hard-soft boundary, and eccentricities are drawn from a thermal distribution \cite{Heggie1975}.

Single and binary stars are evolved with the \textsc{SSE} and \textsc{BSE} codes~\cite{HurleyPols2000,HurleyTout2002}, with up-to-date prescriptions for neutron star and BH formation \cite{FryerBelczynski2012,BelczynskiHeger2016}. Natal kicks for neutron stars born in core-collapse supernovae are sampled from a Maxwellian distribution with a standard deviation of $265\,\rm{km \,s^{-1}}$~\cite{HobbsLorimer2005}, while BH kicks are assumed to be reduced in magnitude according to the fractional mass of the supernova fallback material~\cite{FryerBelczynski2012}.

Each simulation is evolved up to a final time of $14$\,Gyr, unless the cluster disrupts or undergoes a collisional runaway process. The \textsc{cmc} cluster catalog model reproduces well the observed properties of the population of the Milky Way's GCs, including their masses, densities, and ages \citep{Kremer:2019iul}.

\subsection{Globular cluster population}
\label{subsec:GCP}
We characterize the GC population with a differential rate density across cluster masses ($M$), virial radii ($r_v$), metallicities ($Z$), and galactocentric radius ($R_{gc}$) as a function of their formation redshift ($z_{f}$). We make the assumption that the formation mass, virial radius, galactocentric radius, and redshift are all independently distributed, while the distribution of metallicities is contingent upon the redshift
\begin{align}
\label{eq:one}
\frac{\mathrm{d} N_{\mathrm{GC}}}{\mathrm{d} V_c \mathrm{d} t \mathrm{d} M \mathrm{d} r_v \mathrm{d} Z \mathrm{d} R_{gc}} \left(z_f\right) =\, & \mathcal{R}_{\mathrm{GC}}(z_{f}) \; p(Z \mid z_{f} ) \; p(M) \nonumber \\
& \times \; p(r_v) \; p(R_{gc} ),
\end{align}
where $V_c$ is comoving volume and $t$ is the time measured in the
source frame. At a given merger redshift $z_m$, each \textsc{cmc} simulation conducted at a particular gridpoint within the mass–virial radius–metallicity-galactocentric radius parameter space ($M^i, r_v^i, Z^i, R_{gc}^i$) with grid spacing ($\Delta_M^i, \Delta_r^i, \Delta_Z^i, \Delta_R^i$), yields the rate of BBH mergers in the detector frame as
\begin{align}
\label{eq:two}
\mathcal{R}_{\mathrm{dyn}} & \left(z_m \right) = \sum_i p(M^i) \Delta_M^i \; p(r_v^i) \Delta_r^i \; p(R_{gc}^i) \Delta_R^i \nonumber \\
& \times \sum_{l=1}^{N(i)} \frac{1}{1+z_{f,l}} \mathcal{R}_{\mathrm{GC}}\left(z_{f,l}\right) p\left(Z^i | z_{f,l} \right) \Delta_Z^i,
\end{align}
where index $i$ sums over all grid points in the catalog, while index $l$ sums over $N(i)$ BBH merger events in the catalog at the grid point $i$. In the previous equation, $z_{f,l}$ is the formation redshift of the GC within which the $l$th BBH merger event occurs, and it can be determined through the time-to-redshift transformation $z_{f,l}=\hat{z}(t_m + \tau_l)$, where $\tau_l$ is the time that $l$th BBH from the $i$th simulated globular cluster takes to merge. To characterize the cluster formation rate number density $\mathcal{R}_{\mathrm{GC}}(z) = \frac{d N_{\mathrm{GC}}}{d V_c dt} (z)$, a phenomenological Madau-like function~\cite{Madau:2014bja} is adopted
\begin{equation}
\mathcal{R}_{\mathrm{GC}}(z)=\mathcal{R}_0 \frac{(1+z)^{a}}{1+\left[(1+z) /\left(1+z_{\text {peak }}\right)\right]^{a+b}}.
\label{eq:R_GC}
\end{equation}
While it may be tempting to utilize the present-day cluster number density as a direct indicator of $\mathcal{R}_0$ through integrating $\mathcal{R}_{\mathrm{GC}}(z)$ when $a,b$ and $z_{\mathrm{peak}}$ are given, it is important to consider that a significant portion of clusters do not endure until the present time due to evaporation effects~\cite{Rodriguez:2018rmd,Fragione:2018vty,Choksi:2018jnq,Antonini:2019ulv}. Consequently, it becomes necessary to introduce an additional factor to account for evaporation, denoted as $f_{\mathrm{ev}} = \frac{n_0}{n_{\mathrm{surv}}}$. Here, $n_0$ represents the total integral of $\mathcal{R}_{\mathrm{GC}}(z)$, while $n_{\mathrm{surv}}$ stands for the current cluster number density, which we set to $n_{\mathrm{surv}} = 2.31 \times 10^9~ \mathrm{Gpc}^{-3}$~\cite{PortegiesZwart:1999nm}.
We expect that the range of $f_{\mathrm{ev}}$ is limited to $\log_{10} f_{\mathrm{ev}} \in (0,3)$ (see \cite{PortegiesZwart:1999nm, Antonini:2019ulv,Kremer:2019iul,Antonini:2022vib} for details).
For the probability distributions of $p(M), p(r_v)$, $p(R_{gc})$ and $p(Z | z_f)$, we will follow the methodology provided by~\cite{Fishbach:2023xws}. A concise summary of their approach is presented below. 

\begin{figure}[h]
    \centering
    \includegraphics[width=.49\textwidth,left]{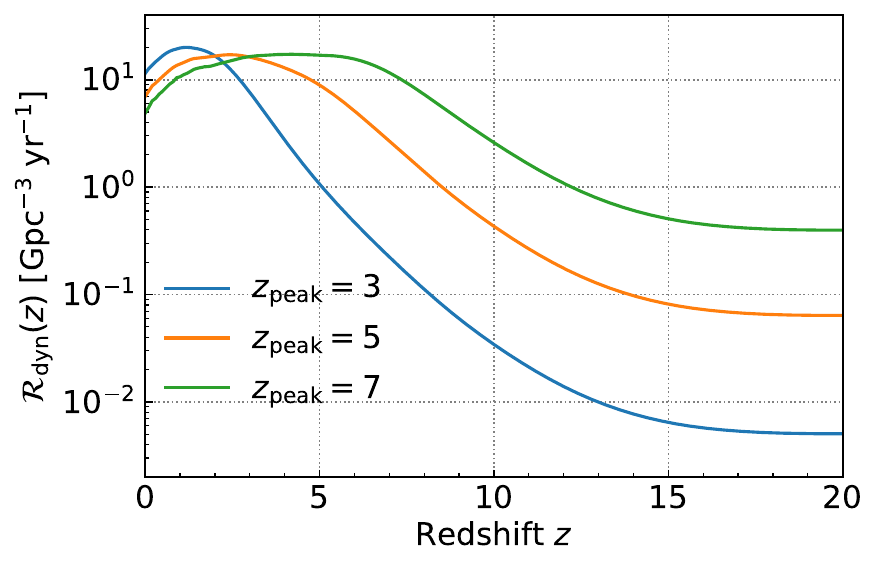}
    \caption{Predicted merger rates in the detector frame for dynamically-assembled BBH for different values of $z_{\mathrm{peak}}$ in Eq.~\ref{eq:R_GC}. These BBH merger rate predictions assume the formation rate density of GCs in the mass range $10^4 - 10^8 M_{\odot}$, with $f_{\mathrm{ev}} = 80$, $\mu_r = 2\;\mathrm{pc}$, $\sigma_r = 1.5\;\mathrm{pc}$, $a=3$ and $b=5$.}
    \label{fig:fig1}
\end{figure}

At each formation redshift, we employ a log-normal distribution of metallicities for the characterization of GCs, where the central metallicity is determined by equation (6) of \cite{Madau:2016jbv} and a metallicity spread of 0.5 dex is assumed. Concerning the mass distribution of GCs, we adopt a Schechter-like function that spans a range from a minimum mass of $10^4 M_{\odot}$ to a maximum mass of $10^8 M_{\odot}$ compatible with GCs observations~\cite{harris2013catalog}. This distribution is characterized by a power-law slope denoted as $\beta_m$ and a Schechter mass denoted as $M^{\star}$
\begin{equation}
p(M) \propto\left(\frac{M}{M^{\star}}\right)^{\beta_m} \exp \left(-\frac{M}{M^{\star}}\right).
\label{eq:four}
\end{equation}
In our study, we maintain fixed values for $\beta_m$ and $M^{\star}$, specifically $\beta_m = -2$ and $M^{\star} = 10^{6.3} M_{\odot}$. It is worth noting that our choice of the mass range is much broader than the range of GC masses in the CMC catalog, which is limited to $1.2 - 19.2 \times 10^5 M_{\odot}$. However, we extend our analysis beyond this range by implementing an extrapolation which assumes that, at a fixed radius, the number of mergers scales with the initial GC mass as $M^{1.6}$ (see \cite{Kremer:2019iul, Antonini:2019ulv, Antonini:2020xnd} for details). The potential impacts of varying the parameters $\beta_m$ and $M^{\star}$ in Eq.~\ref{eq:four} on the SGWB will be left for future investigations. 

For the distribution of galactocentric radii, we allocate equal weights to different values of $R_{\mathrm{gc}}$ for simplicity, considering that the distribution of initial positions of GCs are not yet well-constrained (but see \cite{Gnedin:2013cda} for a recent exploration on this topic). Regarding the distribution of virial radii, we adopt a Gaussian distribution with a mean value of $\mu_r$ and a standard deviation of $\sigma_r$ with a truncation between $0.5\;\mathrm{pc}$ and $4\;\mathrm{pc}$. In our study, we will allow $\mu_r$ to vary from $0.5\;\mathrm{pc}$ to $4\;\mathrm{pc}$ but hold $\sigma_r$ fixed at $1.5\;\mathrm{pc}$ unless specifically stated otherwise.

In Figure~\ref{fig:fig1}, we present the BBH merger rate density in the detector frame as a function of redshift for various $z_{\mathrm{peak}}$ values, while keeping other hyperparameters fixed. It is crucial to observe that higher values of $z_{\mathrm{peak}}$ generally result in a smaller $\mathcal{R}_{\mathrm{dyn}}(0)$; however, this also leads to a higher occurrence of BBH merger events at larger redshift epochs. Consequently, such behavior could have distinct effects on both the shape and amplitude of the energy spectrum of the SGWB, which involves the integration of the GW radiation from all BBH merger events throughout the entire universe.

\subsection{Stochastic GW Background}
\label{sec:sgwb}
The stochastic gravitational-wave background is conventionally described by a dimensionless energy-density spectrum~\cite{Allen:1997ad}
\begin{equation}
\Omega_{\mathrm{GW}}(f)=\frac{1}{\rho_{c, 0}} \frac{\mathrm{d} \rho_{\mathrm{GW}}(f)}{\mathrm{d} \ln f},
\label{eq:five}
\end{equation}
where $\mathrm{d} \rho_{\mathrm{GW}}$ is the present-day gravitational-wave energy density between $f$ and $f + \mathrm{d}f$ and $\rho_{c, 0}$ is the critical energy density in order to close the universe.

For the case of binary coalescences, the energy spectrum contributed by these sources can be written as \cite{Allen:1996vm,Phinney:2001di,Romano:2016dpx}

\begin{align}
\label{eq:six}
\Omega_{\mathrm{GW}}(f) = \frac{f}{c \rho_{c,0}} & \int d\theta  \int dz  \frac{\mathrm{d} V_c}{\mathrm{d} z} \frac{\mathcal{R}_{\mathrm{dyn}}(z)p(\theta,z)(1+z)^2}{4 \pi d_L^2} \nonumber \\
& \times\left. \frac{\mathrm{d} E_{\mathrm{GW}}(\theta)}{d f_r}\right|_{f_r=(1+z) f},
\end{align}
where the integrand of $\Omega_{\mathrm{GW}}(f)$ can be divided into three different components. For the first term, the dependence of the co-moving volume on redshift is given by
\begin{equation}
\frac{\mathrm{d} V_c}{\mathrm{d} z} = \frac{4 \pi c}{H_0} d_M^2 \frac{1}{E\left(\Omega_M,\Omega_\Lambda, z\right)},
\label{eq:eight}
\end{equation}
where $d_M = d_L/(1+z)$ is the proper distance, $d_L$ is the luminosity distance and we set $H_0 = 67.7~ \mathrm{km/s/Mpc}$, $\Omega_{M} = 0.31$ and $\Omega_\Lambda = 1 - \Omega_{M}$ with with $E(\Omega_M, \Omega_\Lambda, z) = \sqrt{\Omega_M (1+z)^3 + \Omega_\Lambda}$ by assuming a $\Lambda\mathrm{CDM}$ cosmology from Planck 2018~\cite{Planck:2018vyg}. The second term $R_{\mathrm{dyn}}(z)p(\theta,z)$ in Eq.~\ref{eq:six} is related to astrophysical details. Here, $\theta$ denotes the source parameters such as mass components of the binary black holes, $\mathcal{R}_{\mathrm{dyn}}(z)$ is the total rate of BBH merger events per comoving volume ($\mathrm{Gpc}^{-3} \mathrm{yr}^{-1}$) at redshift $z$ in the detector frame, and $p(\theta,z)$ is the probability density of source parameters at given redshift. The term $\frac{\mathrm{d} E_{\mathrm{GW}}(\theta)}{d f_r}$ corresponds to the GW energy spectrum emitted by the BBH system evaluated in the source frame per frequency bin. Assuming the BBH systems are in quasi-circular orbits when they reach the LIGO band, we adopt the waveform approximation from \cite{Ajith:2009bn} to compute it\footnote{In this analysis, we have ignored the effect of the BH spin, which does not make a significant difference in the
GW spectrum at low frequencies~\cite{Zhu:2011bd}.} (see Appendix~\ref{sec:waveform} for details).

In Section.~\ref{subsec:GCP}, we explained the procedure to compute the merger rate through a discrete summation by Eq.~\ref{eq:two} starting from the given \textsc{cmc} catalog. For the calculation of $\Omega_{\mathrm{GW}}(f)$ given by Eq.~\ref{eq:six}, we can adopt a similar approach but insert additional terms to GW emission and propagation. We explicitly denote dependence of the emitted energy spectrum on the two BH masses  of the $l$th BBH merger event originating from the $i$th globular cluster class at redshift $z_{m,l}$ as $\mathrm{d} E_{\mathrm{GW}}^l (m_{1,l},m_{2,l})/ \mathrm{d} f_r$. Pulling everything together, we obtain

\begin{widetext}
\begin{equation}
\begin{aligned}
\Omega_{\mathrm{GW}}(f) =& \frac{f}{H_0 \rho_{c,0}} \sum_k \Delta_z^k \sum_i p(M^i) \Delta_M^i \; p(r_v^i) \Delta_r^i \; p(R_{gc}^i) \Delta_R^i \\
& \times \sum_{l=1}^{N(i)} \frac{1}{1+z_{f,l}} \mathcal{R}_{\mathrm{GC}}\left(z_{f,l}\right)\left.\frac{\mathrm{d} E_{\mathrm{GW}}^l(m_{1,l},m_{2,l})}{\mathrm{d} f_r}\right|_{f_r=(1+z_{k}) f} \frac{1}{E\left(\Omega_M,\Omega_\Lambda, z_{k}\right)} \; p(Z^i | z_{f,l}) \Delta_Z^i.
\end{aligned}
\label{eq:cmc_omega}
\end{equation}
\end{widetext}
where the $z_{f,l}$ follows the same time-to-redshift transformation in Eq. \ref{eq:two}, while another summation over the merger redshifts $z_k$, with the redshift grid spacing $\Delta_z^k$ has also been included.

We note that the integration over the source parameters $\theta$ in Eq. \ref{eq:six} is naturally accomplished by summing over all BBH mergers in the catalog. This also automatically includes the variation of the BBH mass distribution with redshift, through the time-delay $\tau_l$ that each BBH takes to merge, as determined by the catalog simulations. In contrast, one can assume a fixed mass distribution for all redshifts (e.g., the power-law-plus-peak (PLPP) phenomenological model~\cite{Talbot:2018cva,KAGRA:2021duu}), which will have a nontrivial impact on the shape of $\Omega_{\mathrm{GW}}(f)$, as we will demonstrate in the next section.

\section{Results}
\label{sec:result}
\subsection{The SGWB Spectrum}
\label{subsec:SGWB}

As described in Section~\ref{subsec:GCP}, our model is characterized by five free parameters: $\boldsymbol{\Theta} = (a, b, z_{\mathrm{peak}}, f_{\mathrm{ev}}, \mu_r)$. While it would be ideal to fully sample all parameters to investigate their effects on $\Omega_{\mathrm{GW}}(f)$, covering all possible combinations becomes computationally infeasible, especially when parameters exhibit a degeneracy with each other (see Appendix~\ref{sec:para} where we discuss in more detail the dependence of $\Omega_{\mathrm{GW}}(f)$ on the parameters). To get a sense of the spectrum $\Omega_{\mathrm{GW}}(f)$ of our model, we start by considering several simplifying assumptions.

First, note that the parameter $b$ primarily governs the merger rate of BBHs at redshifts greater than $z_{\mathrm{peak}}$. This distant part of the BBH population has the smallest contribution to $\Omega_{\mathrm{GW}}(f)$, so given the limited sensitivity of the current and upcoming GW detectors, it is challenging to observe the impact from varying $b$~\cite{Callister:2020arv}. Hence, we will always keep $b$ fixed at a fiducial value $b = 5$, unless specifically indicated otherwise. Second, $f_{\mathrm{ev}}$ is a scaling parameter so it only affects the overall amplitude of $\Omega_{\mathrm{GW}}(f)$ rather than its frequency dependence. The same is true for $\mu_r$, as demonstrated in more detail in Appendix~\ref{sec:para}. These two parameters are, therefore, degenerate, and can be replaced by a single scaling parameter. And third, the remaining parameters ($a$, $z_{\mathrm{peak}}$) primarily control the shape of $\Omega_{\mathrm{GW}}$. 
Since our primary interest is in the importance of $z_{\mathrm{peak}}$, we will fix $a=3$ for most of our calculations. However, we also provide an analysis of different $a$ values in Appendix~\ref{sec:para}.\footnote{A joint analysis of $(a, z_{\mathrm{peak}})$ of the SGWB combined with individual CBC events is presented in \cite{Callister:2020arv}, assuming a redshift-independent BBH mass distribution.} 

\begin{figure}[t]
\includegraphics[width=.49\textwidth,left]{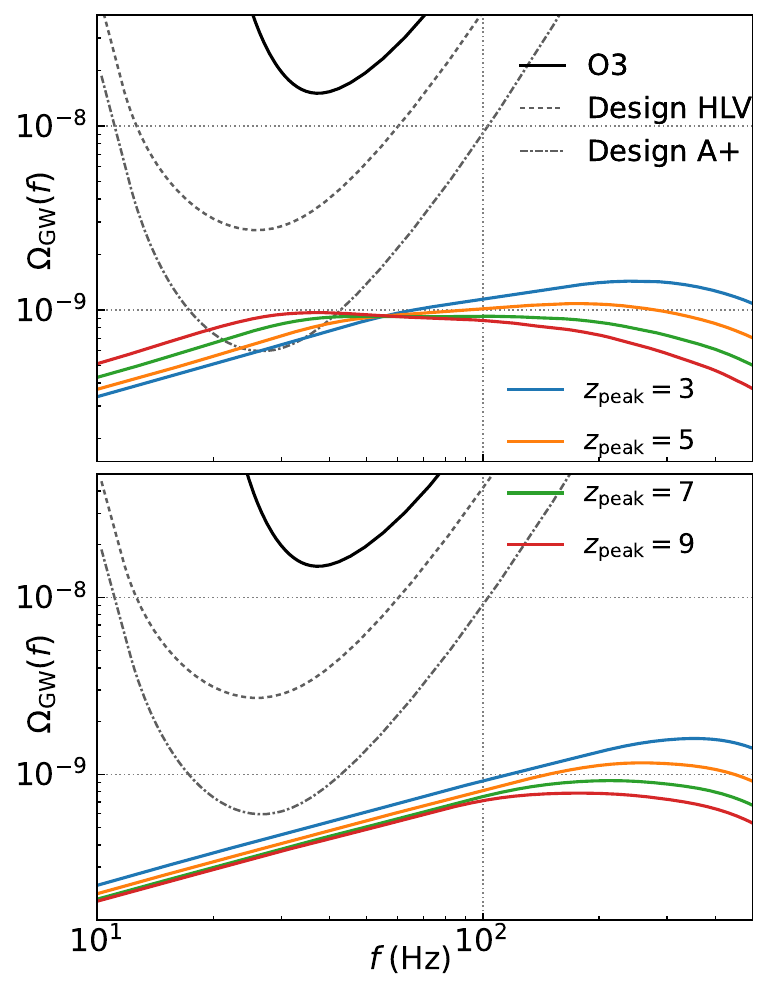}
\caption{Top: $\Omega_{\mathrm{GW}}(f)$ from the \textsc{cmc} model is shown for different values of the $z_{\mathrm{peak}}$ parameter, after fixing $a=3$, $b=5$, $f_{\mathrm{ev}}=80$, and $\mu_r=2\mathrm{\; pc}$. Black lines give $2\sigma$ power-law integrated (PI) curves for O3, projections for the Advanced HLV network at design sensitivity, and A+ detectors. Bottom: same as the top panel, but using the PLPP mass distribution instead of the BBH masses given by the \textsc{cmc} catalog.}
\label{fig:fig2}
\end{figure}

With the above assumptions, the SGWB spectra calculated using Eq.~\ref{eq:cmc_omega} are shown in Figure~\ref{fig:fig2} (top). 
We fix parameters $a=3$, $b=5$, $f_{\mathrm{ev}}=80$, and $\mu_r=2\mathrm{\; pc}$, but vary the value of $z_{\mathrm{peak}}$, showing that both the spectral shape and amplitude of $\Omega_{\rm GW}(f)$ change as a function of $z_{\mathrm{peak}}$. We also show the $2\sigma$ power-law integrated (PI) curves~\cite{Thrane:2013oya} which represent the integrated sensitivity of the O3 search~\cite{KAGRA:2021kbb}, along with projections for 2 year observation of the Advanced LVK network at design sensitivity with a 50\% duty cycle, and for the A+ design sensitivity integrated for 18 months~\cite{KAGRA:2013rdx}. The plot demonstrates that at A+ sensitivity, GW detectors will start to discern the value of the $z_{\mathrm{peak}}$ parameter. 

The dependence of $\Omega_{\rm GW}(f)$ on $z_{\mathrm{peak}}$ arises due to two effects: (i) larger $z_{\mathrm{peak}}$ implies that a larger fraction of the BBH mergers is redshifted to lower frequencies, $<50$ Hz; and (ii) the BBH mass distribution at higher redshifts is shifted to larger masses, resulting in stronger GW contributions at lower frequencies. The latter effect has not received much attention in the literature, and in fact it is common to assume the BBH mass distribution to be redshift-independent and defined by the power-law-plus-peak (PLPP) fit to the BBH mergers observed by the LVK detectors at low redshifts~\cite{Callister:2020arv,KAGRA:2021kbb,Lehoucq:2023zlt,Turbang:2023tjk}. To highlight the importance of the redshift-dependent BBH mass distribution, we will consider two models: the \textsc{cmc} model will use the distribution of BBH masses as determined in the \textsc{cmc} catalog, while the PLPP model will use the same BBH merger rate density from the \textsc{cmc} catalog but it will replace the BBH masses with those drawn from the (redshift-independent) PLPP fit.

\begin{figure}[t]
    \centering
    \includegraphics[width=.49\textwidth,left]{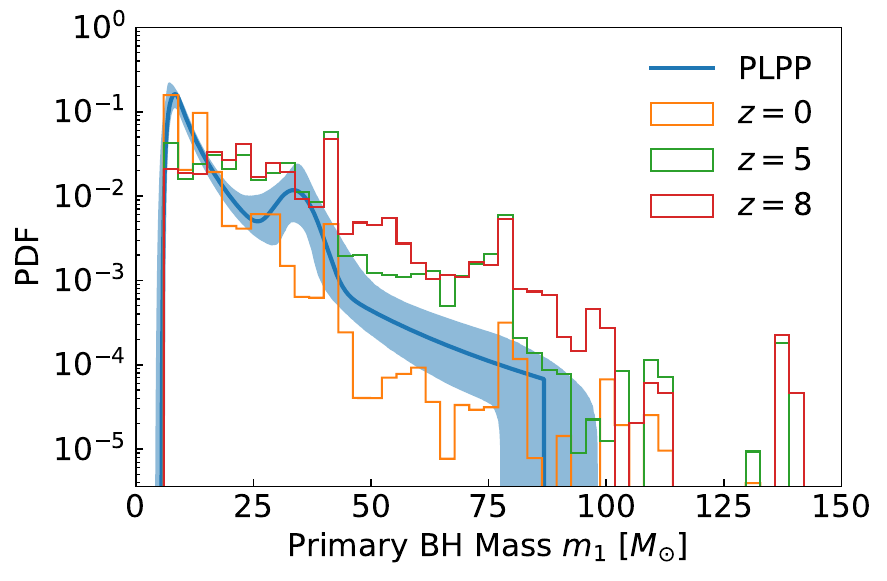}
    \caption{Probability distribution function of primary BH masses ($m_1$) of BBHs merging at redshift  $z=0$ (orange), $z = 5$ (green) and $z=8$ (red) given $z_{\mathrm{peak}} = 5$. The blue solid line is the median value of the PLPP model applied to GWTC-3 BBHs. The shaded blue areas are the corresponding 90\% credible intervals.}
    \label{fig:fig3}
\end{figure}

Figure~\ref{fig:fig3} shows the distribution of masses in the \textsc{cmc} model at different redshifts (assuming $z_{\mathrm{peak}} = 5$), in comparison with the PLPP distribution \footnote{The full list of parameters and their corresponding best-fit values can be found in \cite{KAGRA:2021duu}, with a specific choice of $M_{\textrm{max}}\sim 100\,M_{\odot}$ serving as the cut-off.}. At small redshifts, the primary black hole masses in the \textsc{cmc} model can be reasonably described by the PLPP fit within the 90\% credible intervals of the GWTC-3 catalog, though this fitting is not applicable for BHs with masses exceeding $100\,M_{\odot}$. On the other hand, at larger redshifts ($z > 5$), the \textsc{cmc} mass distribution tends to shift towards higher mass regions and the PLPP model no longer describes the mass distribution based on the \textsc{cmc} catalog.

The implications of the BBH mass distribution are highlighted in Figure~\ref{fig:fig2} (bottom), where we used Eq.~\ref{eq:cmc_omega} with the same \textsc{cmc} catalog, but we replaced the BBH masses with those drawn from the (redshift-independent) PLPP distribution. Again we fix $a=3$, $b=5$, $f_{\mathrm{ev}}=80$, and $\mu_r=2\mathrm{\; pc}$. We observe that increasing $z_{\mathrm{peak}}$ changes the high-frequency behavior of the spectrum, but the low-frequency region (below 100 Hz) is essentially altered by an overall scaling of the amplitude with holding a similar shape at the most sensitive A+ design bands. Additionally, we find that the relationship between the scaling values and the $z_{\mathrm{peak}}$ parameters exhibits an opposite trend compared to the \textsc{cmc} model.

The stark difference between the top and bottom panels of Figure~\ref{fig:fig2} shows that the SGWB measurement with the upcoming terrestrial detectors, in the frequency band between 20 Hz and $\sim 60$ Hz, will be sensitive to the redshift evolution of the BBH mass distribution. Since different channels of BBH formation (field channels vs dynamical channels) may lead to different BBH mass distributions (and their redshift evolutions), these SGWB measurements could therefore be used to probe the BBH formation channels and potentially differentiate between different BBH populations. We will further investigate this effect by performing more detailed parameter estimation in Sections~\ref{subsec:constraints} and \ref{subsec:injection}. 
We also note that below 20 Hz, $\Omega_{\mathrm{GW}}(f) \propto f^{2/3}$ for both \textsc{cmc} and PLPP models and across a range of $z_{\mathrm{peak}}$ values. Hence, such low frequencies will likely not help in studying the mass-redshift relation, although they will be important for inferring the amplitude dependence on $f_{\mathrm{ev}}$. Similarly, at frequencies above $\sim 60$ Hz the weaker SGWB sensitivities will not support such inference studies.

\subsection{Constraints on GC population parameter space}
\label{subsec:constraints}

We next study the accessibility of the model described above to the current and upcoming terrestrial GW detectors. The most recent measurement of the SGWB energy density spectrum, $Y_{\rm GW}(f)$, and the corresponding uncertainty, $\sigma_{\rm GW}(f)$, were obtained using Advanced LIGO and Advanced Virgo data from the first three observing runs~\cite{KAGRA:2021kbb}. To study the implications of this result for our model, we use a Bayesian parameter estimation formalism. Bayes' Theorem defines the posterior distribution of parameters $p(\boldsymbol{\Theta}|Y_{\rm GW})$ as
\begin{eqnarray}
    p(\boldsymbol{\Theta}|Y_{\rm GW}) \propto \mathcal{L}(Y_{\rm GW}|\boldsymbol{\Theta}) \Pi(\boldsymbol{\Theta}),
    \label{eq:posterior}
\end{eqnarray}
where $\Pi(\boldsymbol{\Theta})$ represents the prior distribution of the parameters $\boldsymbol{\Theta}$ and the likelihood is given by
\begin{align}
\label{eq:gwprob}
\mathcal{L} & \left(Y_{\mathrm{GW}} (f) \mid \boldsymbol{\Theta}\right) \propto \nonumber \\
& \exp \left[-\frac{1}{2} \sum_f \left(\frac{Y_{\mathrm{GW}}(f) - \Omega_{\mathrm{GW}}(f \mid \boldsymbol{\Theta}) }{\sigma_{\mathrm{GW}}(f)}\right)^2\right].
\end{align}

\begin{figure}
\centering
\includegraphics[scale=0.52]{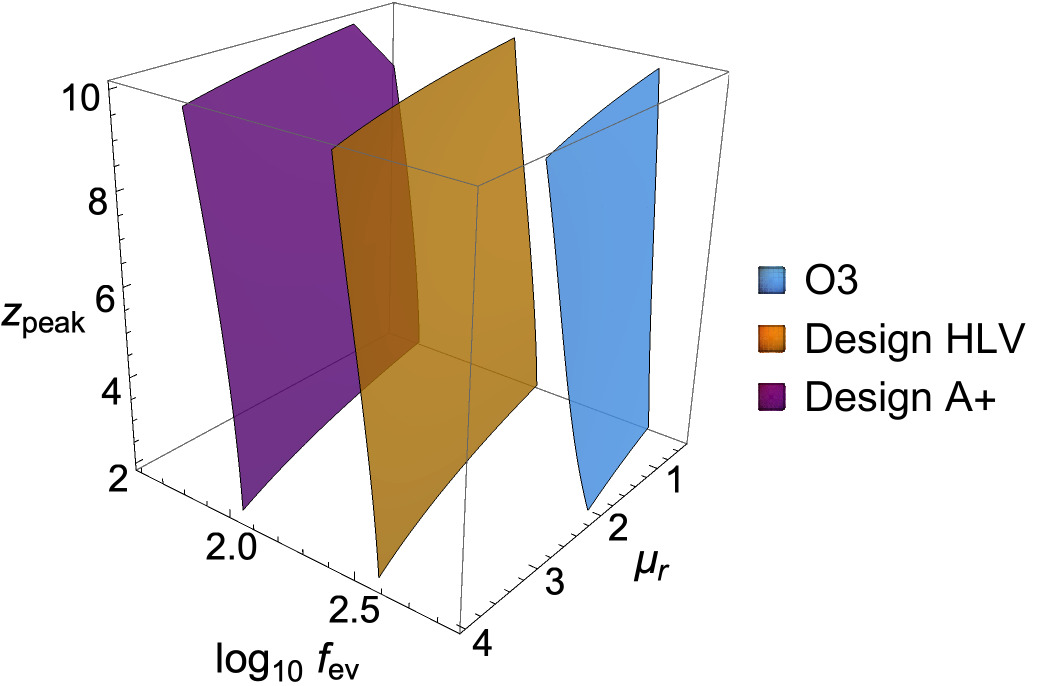}
\includegraphics[width=.48\textwidth,center]{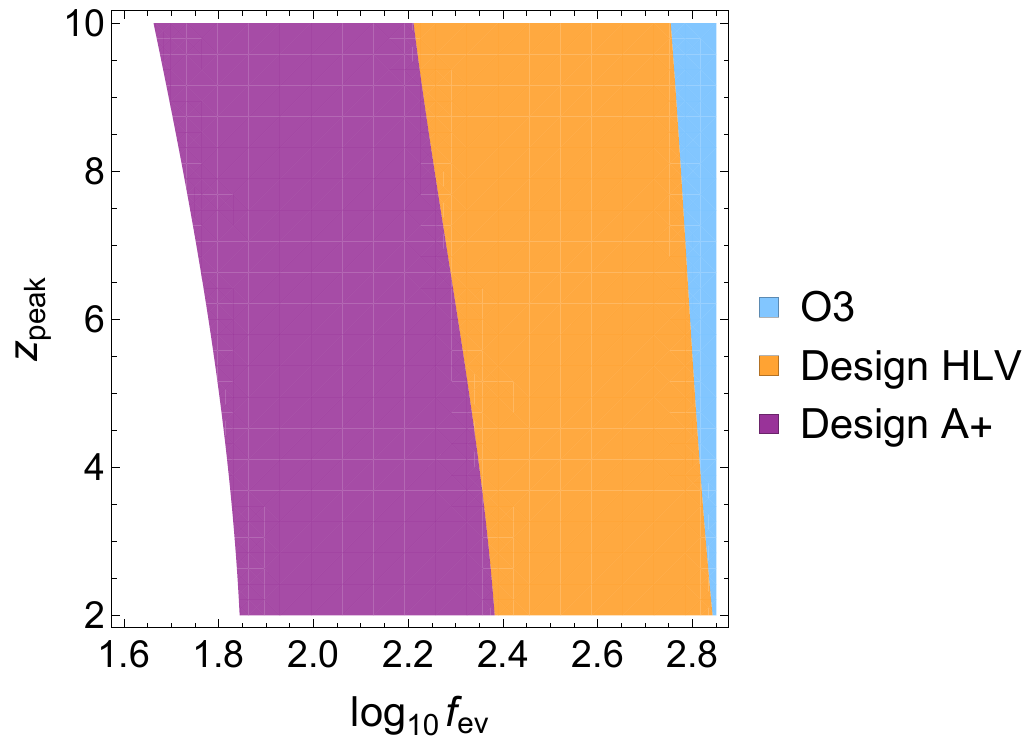}
\includegraphics[width=.48\textwidth,center]{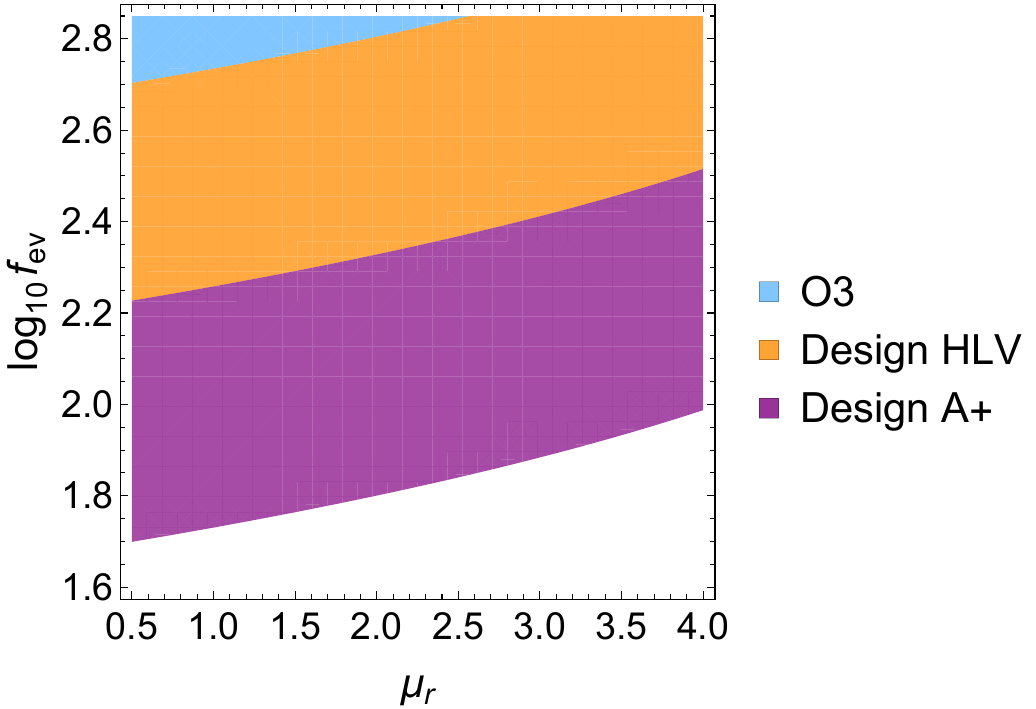}
\caption{Top: Constraints in the GC model parameter space ($z_{\mathrm{peak}},\log_{10}f_{\mathrm{ev}}, \mu_r$) are shown based on the non-observation of SGWB by Advanced LIGO and Advanced Virgo first three observation runs. Also shown are projected sensitivities of the SGWB searches using the design strain sensitivity of Advanced LIGO and Advanced Virgo detectors and using the design A+ sensitivity. The regions above each surface are excluded or accessible to the corresponding measurement at the $95\%$ credible level for the prior choice explained in the text. 
Middle: Slicing of the top figure in the $z_{\mathrm{peak}} - \log_{10}f_{\mathrm{ev}}$ plane after fixing $\mu_r = 2.0 ~\mathrm{pc}$. Bottom: Slicing of the top figure in the $\log_{10}f_{\mathrm{ev}} - \mu_r$ plane after fixing $z_{\mathrm{peak}} = 5$.
}
\label{fig:fig5}
\end{figure}

We fix the values of $a=3$ and $b=5$ to focus on the parameter subspace defined by ($f_{\mathrm{ev}}, \mu_r, z_{\mathrm{peak}}$). Regarding the choice of the prior $\Pi(\boldsymbol{\Theta})$ when performing the Bayesian inference, we apply a uniform prior distribution to $z_{\mathrm{peak}}$ and $\mu_r$ within the intervals of $(2,10)$ and $(0.3,4.8)$ respectively. However, for $f_{\mathrm{ev}}$, a log-uniform distribution is employed, spanning the range between 0 and 3. We also investigate the sensitivity of upcoming SGWB searches using the design strain sensitivity of the Advanced LIGO and Advanced Virgo detectors with 2 years exposure at 50\% duty cycle, and using A+ sensitivity with 18 months exposure~\cite{KAGRA:2013rdx}. To do so, we again use the formalism of Eq. \ref{eq:posterior}, setting $Y_{\rm GW}=0$ and defining the uncertainty as
\begin{equation}
\sigma_{\mathrm{GW}}^{I J} (f) =\sqrt{\frac{1}{ T \Delta f} \frac{P_{I}(f) P_{J}(f)}{\gamma_{I J}^2(f) S_0^2(f)}},
\end{equation}
where $P_{I} (f)$ is the expected noise power spectral density (PSD) of GW detector $I$ at a frequency bin $f$, $\gamma_{I J}(f)$ is the normalized isotropic overlap reduction function quantifying the geometrical sensitivity of a given detector pair to an isotropic gravitational-wave background~\cite{Christensen:1992wi,Flanagan:1993ix}, $T$ is the total observation time, $\Delta f$ is the frequency resolution (we choose $\Delta f = 1/32 \mathrm{Hz}$) and the function $S_0(f)=\frac{3 H_0^2}{10 \pi^2} \frac{1}{f^3}$ converts GW strain power spectrum into the fractional energy density.
We follow the approach of~\cite{Callister:2016ewt} and scale the uncertainty by a factor of $\sqrt{2}$ so as to account for marginalization over the future detector strain noise realizations.

\begin{figure}
    \centering
    \includegraphics[width=.48\textwidth,left]{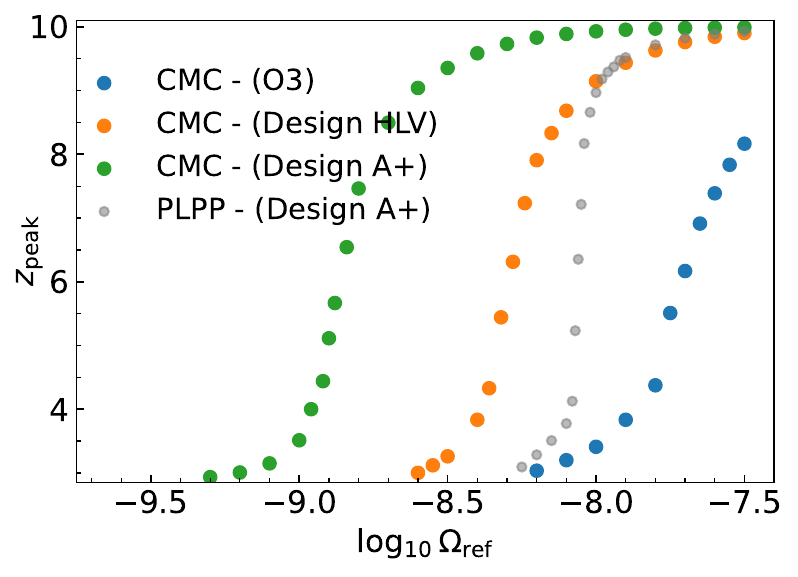}
    \caption{95\% confidence exclusion curve in the $z_{\mathrm{peak}} - \Omega_{\mathrm{ref}}$ parameter space is shown for the O3 observation run and assuming the \textsc{cmc} BBH mass distribution model, along with the design and A+ projected sensitivities. For the A+ sensitivity we also show the curve using the PLPP BBH model. See text for further details.}
    \label{fig:fig8}
\end{figure}

Figure~\ref{fig:fig5} shows the constraints derived in this way in the 3D parameter space of ($\log_{10}f_{\mathrm{ev}}, \mu_r, z_{\mathrm{peak}}$), as well as in 2D slices of the parameter space, where the contour surfaces of the top plot are obtained through evaluating posterior distribution $p(\boldsymbol{\Theta}|Y_{\mathrm{GW}}(f))$ at the $95\%$ credible level. While the existing result from Advanced LIGO and Advanced Virgo is already ruling out a part of the interesting parameter space in this model, the terrestrial GW detector network at the design and A+ sensitivity has the potential to explore a significant portion of the parameter space. Our new constraints on the GC population parameter space complement the work presented in \cite{Fishbach:2023xws}. In that study, the inference on relevant GC parameters is based on BBH merger rate analysis from GWTC-3 catalog, but the ability to impose constraints at large $z_{\mathrm{peak}}$ values is limited (see Figure A2 in \cite{Fishbach:2023xws}) due to the fact that GWTC-3 catalog only contains local merger events, making it challenging to extract information about BBH merger rate history beyond $z\sim3$.

\begin{figure*}[ht]
\centering
\begin{minipage}{\columnwidth}
\includegraphics[width=\textwidth,center]{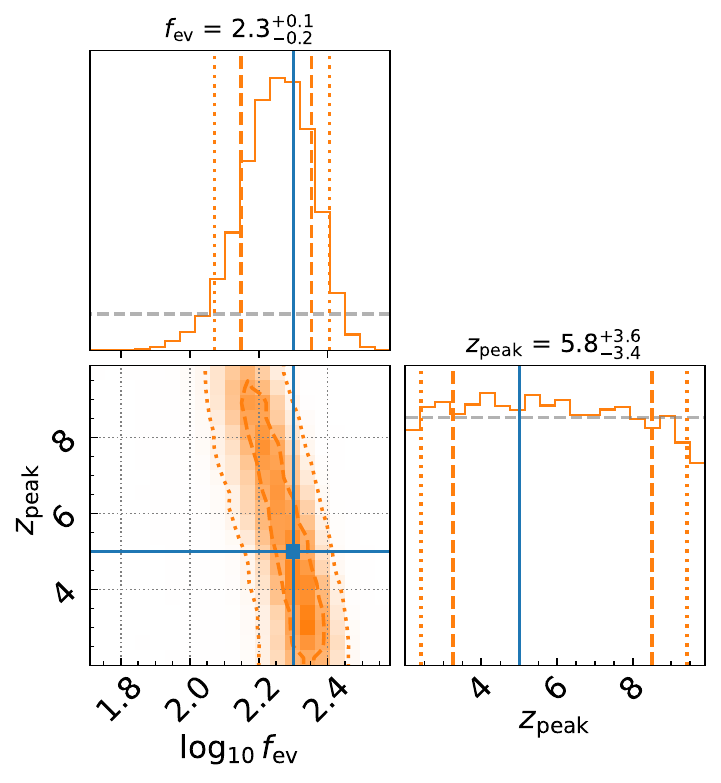}
\end{minipage}
\begin{minipage}{\columnwidth}
\includegraphics[width=\textwidth,center]{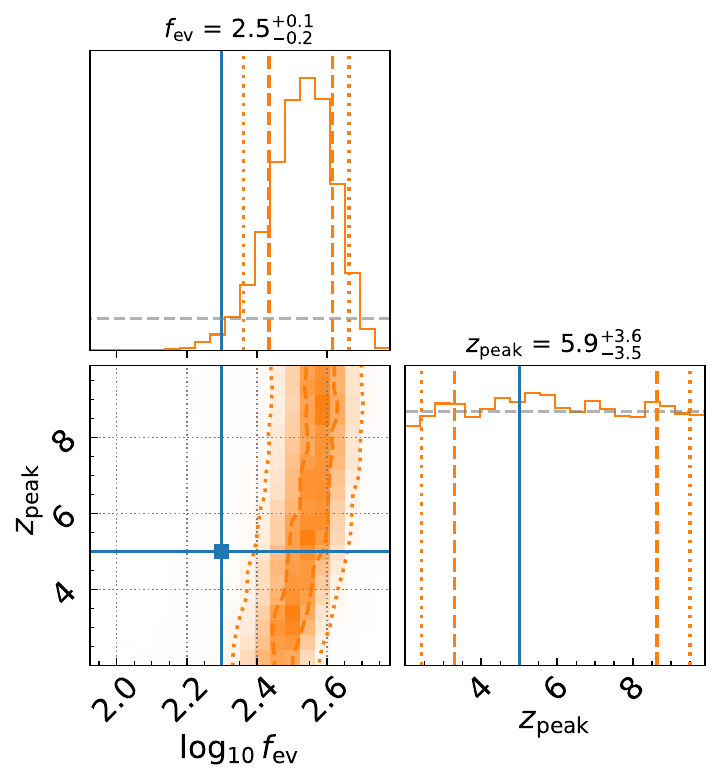}
\end{minipage}
\caption{Left: Corner plot obtained from running the $f_{\mathrm{ev}}, z_{\mathrm{peak}}$ parameter estimation using SGWB based on the \textsc{cmc} model and assuming the design A+ sensitivity, where dashed and dotted lines represent $1\sigma$ and $2\sigma$ contours respectively. The simulated values are represented by the blue lines ($\log_{10}f_{\mathrm{ev}}=2.3$ and $z_{\mathrm{peak}} = 5$). Vertical orange dashed and dotted lines indicate $1\sigma$ and $2\sigma$ confidence intervals, while dashed gray lines illustrate the prior distribution. Right: Same as left panel, but using the PLPP model for BBH mass distribution in the recovery.}
\label{fig:contour_aplus}
\end{figure*}

Finally, as another way to examine the parameter space, we fix two parameters $a=3, b=5$, and vary $z_{\mathrm{peak}}$ over the range $(2,10)$. For each choice of $z_{\mathrm{peak}}$, we adjust the value of the scaling parameter $f_{\mathrm{ev}}$ so as to maintain a constant value of $\Omega_{\rm ref} = \Omega_{\rm GW} (20 {\rm \; Hz})$.
For each value of $\Omega_{\rm ref}$ we determine the value of $z_{\mathrm{peak}}$ consistent with the existing or future measurement at 95\% confidence level, and we show results in Figure~\ref{fig:fig8}. With increasing sensitivity of the detector network, from the O3 observation run to the future A+ upgrade, larger regions of the $z_{\mathrm{peak}} - \Omega_{\mathrm{ref}}$ plane can be explored. In particular, if the amplitude of the SGWB at 20 Hz is $\Omega_{\rm ref} = 10^{-9}$, the experiments will be able to explore the GC peak redshift up to $z_{\mathrm{peak}}\sim 3.5$, and significantly higher if the amplitude is larger. We also note that at any given sensitivity level, e.g. Design A+, utilizing the SGWB spectrum calculated through the \textsc{cmc} model (green points) yields more stringent constraints compared to the PLPP case (gray points). This outcome is not surprising because the peak of the spectrum in the \textsc{cmc} model is better aligned with the most sensitive frequency band of the detectors.

\subsection{Injection and Recovering Study}
\label{subsec:injection}
As another illustration of the importance of correctly modelling the BBH mass distribution and its redshift dependence, we perform a simulation recovery study. In particular, we fix three of the GC population parameters ($a=3$, $b=5$, $\mu_r=2\mathrm{\; pc}$) and consider the remaining two parameters ($z_{\mathrm{peak}}$ and $f_{\mathrm{ev}}$) as free. To simulate the stochastic GW background, we assume $z_{\mathrm{peak}} = 5$, $\log_{10}f_{\mathrm{ev}} = 2.3$, and use BBH masses as provided by the \textsc{cmc} model, which gives a rate of BBH mergers of $15 \;\mathrm{Gpc}^{-3}\,\mathrm{yr}^{-1}$ at $z=0$ and an amplitude of the SGWB $\Omega_{\mathrm{ref}} = 1.4\times 10^{-9}$ at $20$ Hz. 

We then attempt to recover this SGWB using two different models, and assuming the Design A+ sensitivity HLV network observing for 18 months. Our recovery follows the Bayesian parameter estimation approach (Eq. \ref{eq:posterior}), with a flat prior on $\log_{10}f_{\mathrm{ev}}$ within the interval of $(0,3)$~\cite{Fishbach:2023xws} and a uniform prior within the interval of $(2,10)$ for $z_{\mathrm{peak}}$, respectively. 

\begin{figure*}[ht]
\centering
\begin{minipage}{\columnwidth}
\includegraphics[width=\textwidth,center]{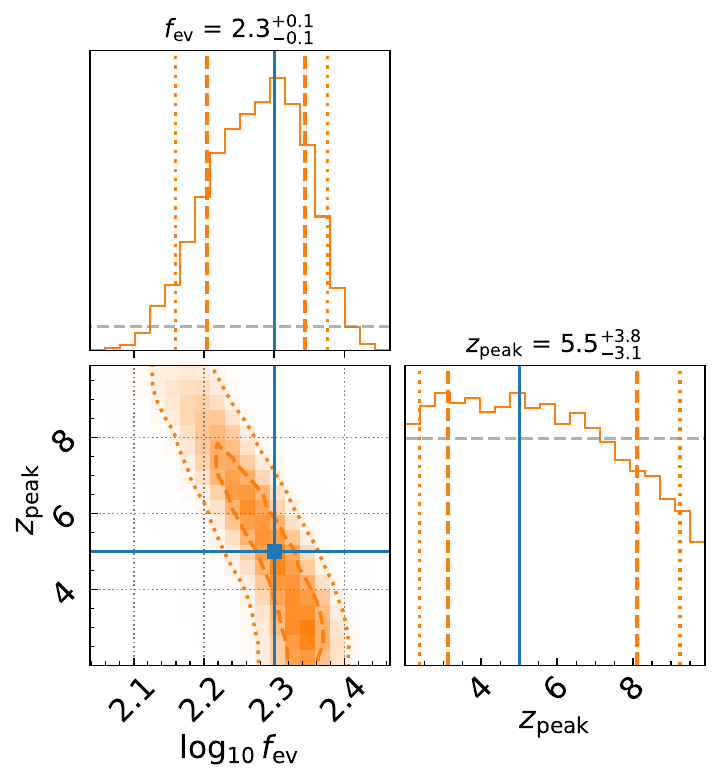}
\end{minipage}
\begin{minipage}{\columnwidth}
\includegraphics[width=\textwidth,center]{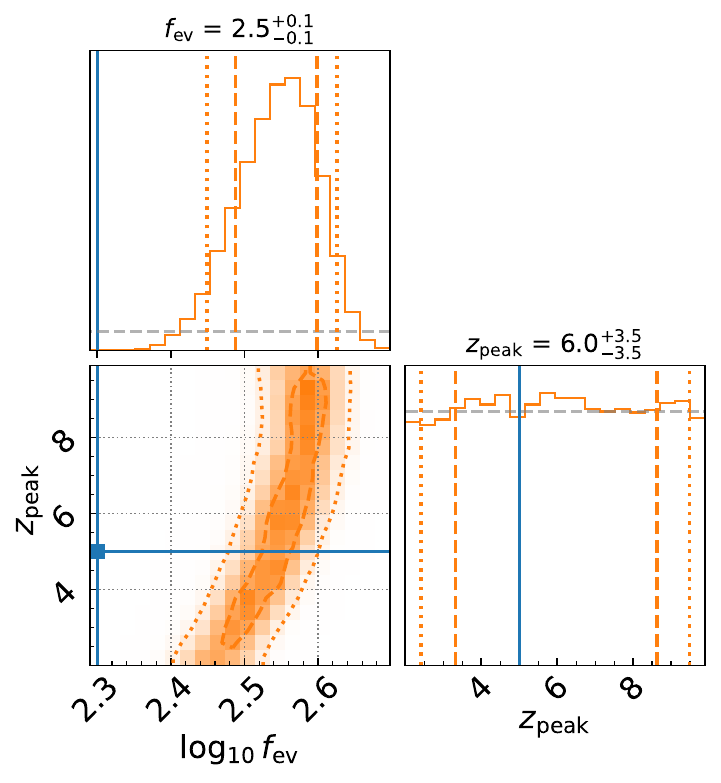}
\end{minipage}
\caption{Left: Corner plot obtained from running the $f_{\mathrm{ev}}, z_{\mathrm{peak}}$ parameter estimation using SGWB based on the \textsc{cmc} model and assuming the design A\# sensitivity, where dashed and dotted lines represent $1\sigma$ and $2\sigma$ contours respectively. The simulated values are represented by the blue lines ($\log_{10}f_{\mathrm{ev}}=2.3$ and $z_{\mathrm{peak}} = 5$). Vertical orange dashed and dotted lines indicate $1\sigma$ and $2\sigma$ confidence intervals, while dashed gray lines illustrate the prior distribution. Right: Same as left panel, but using the PLPP model for BBH mass distribution in the recovery.}
\label{fig:contour_asharp}
\end{figure*}

In the first recovery, we use the \textsc{cmc} model to search the parameter space. The results are shown 
in Figure~\ref{fig:contour_aplus} (left panel). The simulated parameter values are recovered well, as shown in the 2D posterior; however, the individual parameters are not localized well, especially for $z_{\mathrm{peak}}$ parameter, due to their degeneracy. 

In the second recovery, we use the redshift-independent BBH mass distribution given by the PLPP model. Figure ~\ref{fig:contour_aplus} (right panel) shows the results of this recovery. Even though extracting information about $z_{\mathrm{peak}}$ remains equally challenging, this model fails to accurately capture the simulated value of the parameter $f_{\mathrm{ev}}$, preferring large values due to the reversed relationship between $z_{\mathrm{peak}}$ and the overall $\Omega_{\mathrm{GW}}(f)$ amplitude (see Fig~\ref{fig:fig2}).

The situation worsens as the sensitivity of the GW detector network improves. We repeat the simulation with the same parameter values, but now for the recovery we assume even better, A\# detector strain sensitivity~\cite{T2200287,Gupta:2023lga}
with four years observation time in the LIGO-Virgo network. 
As shown in Figure~\ref{fig:contour_asharp} (left panel) with the \textsc{cmc} model, the regions with $z_{\mathrm{peak}}\geq 8$ are disfavoured thanks to a factor of 2 improvements compared to the A+ strain sensitivity~\cite{KAGRA:2013rdx}. However, assuming the PLPP model not only introduces a bias in the estimation of $f_{\mathrm{ev}}$ but also hinders the extraction of information on the $z_{\mathrm{peak}}$ parameter, as shown in Figure~\ref{fig:contour_asharp} (right).

This analysis highlights the importance of incorporating realistic mass distributions when conducting model selection and parameter estimation with the SGWB. In reality, the background might receive contribution from multiple channels of BBH formation, with potentially different evolution of mass components distribution~\cite{Mukherjee:2021ags,Atal:2022zux}. While this makes the SGWB modeling and data analysis more complicated, such complexity seems necessary in order to avoid potential biases in inference studies. In particular, inclusion of redshift dependence of BBH mass distribution appears to be critical. A detailed analysis of the redshift dependent mass distribution by combining multiple BH formation channels is left for future work.

\section{Summary}
\label{sec:summary}
In this work, we conducted an investigation into the SGWB arising from the coalescing BBHs dynamically assembled within dense star clusters, based on the \textsc{cmc} catalog \citep{Kremer:2019iul}. Compared to previous studies, we are not restricted to phenomenological cluster evolution models or oversimplified assumptions about the population of star clusters in the universe. Adopting this more sophisticated approach not only strengthened the robustness of our results but also yielded two additional important benefits:
\begin{enumerate}[label=(\roman*)]
\item {\it Improved Constraints on GC Populations}: By employing a more realistic description of GC populations and considering a broader range of GC models that describe their initial properties, we could assess how different GC populations influence both the amplitudes and the shape of the SGWB spectrum. Consequently, we demonstrated how the SGWB spectrum measurements can be used to constrain the relevant GC population parameters.
\item {\it Evolution of BBH Mass Components}: Our approach allowed us to trace the evolution of the distribution of BBH mass components at different redshifts. We demonstrated that this mass distribution evolution can have a significant effect on the SGWB spectrum, hence providing a way do distinguish between different astrophysical channels for the origin of BBH mergers.
\end{enumerate}

Regarding the first point, we demonstrated that the current SGWB search results already constrain the parameter region ($z_{\mathrm{peak}},f_{\mathrm{ev}},\mu_r$) of GC populations, though a large portion of the parameter space is still allowed. However, we also showed that with future detector upgrades, larger areas of the parameter space will be constrained. 
~Meanwhile, advancements in galaxy simulations and high-redshift electromagnetic observations, particularly those facilitated by the recently launched JWST, are enhancing our understanding of GCs \citep{VanzellaCastellano2022,VanzellaClaeyssens2023}. This evolving knowledge base will offer improved guidance for selecting GC models, especially with regard to the cluster formation rates model adopted in this paper. Our study underscores the utility of SGWB searches in providing complementary insights into the physical characteristics of GCs, complementing high-redshift measurements in the coming years.

As for the second point, our results revealed that the distribution of primary mass components of BBHs at high redshifts in the \textsc{cmc} model deviates from the phenomenological PLPP model. While at lower redshifts the mass distribution can still reasonably match the PLPP model below $\sim 100 M_{\odot}$, at high redshifts the \textsc{cmc} mass distribution shifts towards higher-mass regions. This observation has important implications when computing the energy spectrum $\Omega_{\mathrm{GW}}(f)$: the heavy BBHs at higher redshifts lead to a shift in the peak of $\Omega_{\mathrm{GW}}(f)$ towards lower frequencies, which enhances the accessibility of the SGWB given the detector network's optimal sensitivity around 25 Hz. This specific BBH mass evolution, as opposed to a redshift-independent distribution, enables distinguishing between SGWB models and performing accurate parameter estimation of the SGWB. To the best of our knowledge, our paper is the first to demonstrate this effect, underscoring the necessity of incorporating population synthesis outcomes when conducting SGWB modeling.

We highlight, however, some limitations of our analysis and some extensions of our work that should be investigated in the future. First of all, while we have achieved a significant improvement in modeling the dependence between the GC population and the SGWB, it is essential to recognize that specific modeling choices can still influence our results. While we have studied the impact of parameters on the shape and overall amplitude of $\Omega_{\mathrm{GW}}(f)$, our study has not been exhaustive. For instance, the joint effects of parameters like the low-redshift slope $a$ and the peak redshift $z_{\mathrm{peak}}$ on the SGWB energy spectrum warrant further investigation---as we show in Appendix A, larger $a$ can also move the peak position of $\Omega_{\mathrm{GW}}(f)$ toward lower frequencies. Additionally, exploring different functional forms for the GC formation rate density is essential to obtain a more comprehensive understanding of how different models and assumptions affect the SGWB predictions. We also note that novel search approaches leveraging the intermittent nature of the BBH background could be significantly more sensitive than the traditional search for isotropic SGWB~\cite{Smith:2017vfk,Lawrence:2023buo,Sah:2023bgr,Dey:2023oui}. Such approaches could consequently improve model selection and parameter estimation of BBH models. 

\begin{acknowledgments}
The authors are grateful
for computational resources provided by the LIGO Laboratory and supported by National Science Foundation
Grants PHY-0757058 and PHY-0823459. G.F. acknowledges support by NASA Grant 80NSSC21K1722 and NSF Grant AST-2108624 at Northwestern University. V.M. is in part supported by the NSF grant PHY-2110238 at the University of Minnesota.
This material is based upon work supported by NSF's LIGO Laboratory which is a major facility fully funded by the National Science Foundation.
\end{acknowledgments}

\appendix
\section{Investigations of selected GC population parameters.}
\label{sec:para}
\begin{figure}[h]
\centering
\includegraphics[width=.5\textwidth,left]{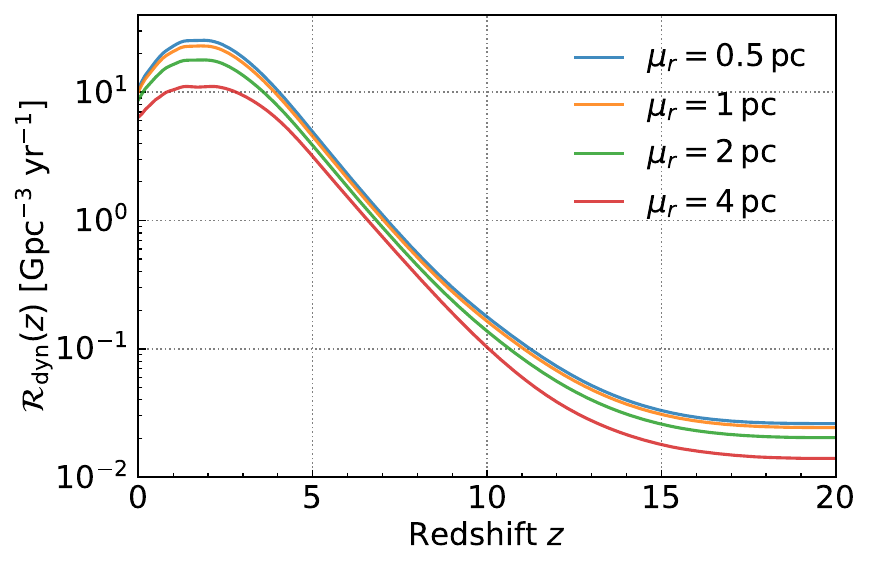}
\caption{Inferred rate of dynamically-assembled BBH mergers in the detector frame with different virial radius distributions parameterized by $\mu_r$ (colored, solid lines). These BBH rate predictions assume that the formation rate density of GCs follows the dashed black line, with $a=3$, $b=5$, $z_{\mathrm{peak}} = 4$, $f_{\mathrm{ev}} = 80$, and $\sigma_r = 1.5\;\mathrm{pc}$.}
\label{fig:fig16}
\end{figure}

\begin{figure}[h]
    \centering
    \includegraphics[width=.475\textwidth,left]{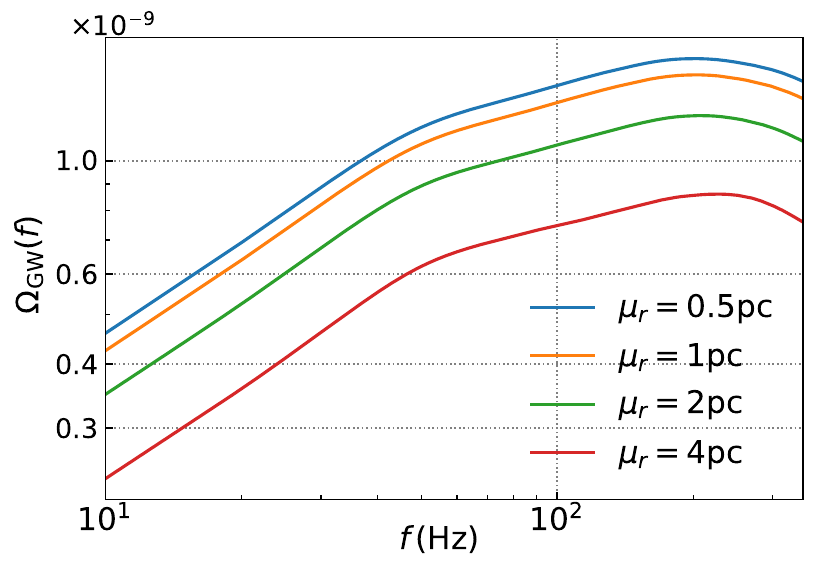}
    \caption{$\Omega_{\mathrm{GW}}(f)$ corresponding to different values of $\mu_r$, while keeping other parameters the same as Figure~\ref{fig:fig16}.}
    \label{fig:fig17}
\end{figure}
In this Appendix we present the effects of another two parameters from GC populations, namely $\mu_r$ and $a$, on the SGWB spectrum. The parameter $\mu_r$ governs the weighting of different GC models, with heavier GCs expected to host a greater number of BBHs thus increasing the BBH merger rate. In Figures~\ref{fig:fig16} and \ref{fig:fig17}, we show the dependence of $\mathcal{R}_{\mathrm{dyn}}(z)$ and $\Omega_{\mathrm{GW}}(f)$ on $\mu_r$, after fixing other parameters. It can be observed that changing the value of $\mu_r$ will only alter the overall amplitude of the spectrum, without changing the shape of the BBH merger rate or the spectrum of the SGWB significantly.

\begin{figure}
\centering
\includegraphics[width=.5\textwidth,left]{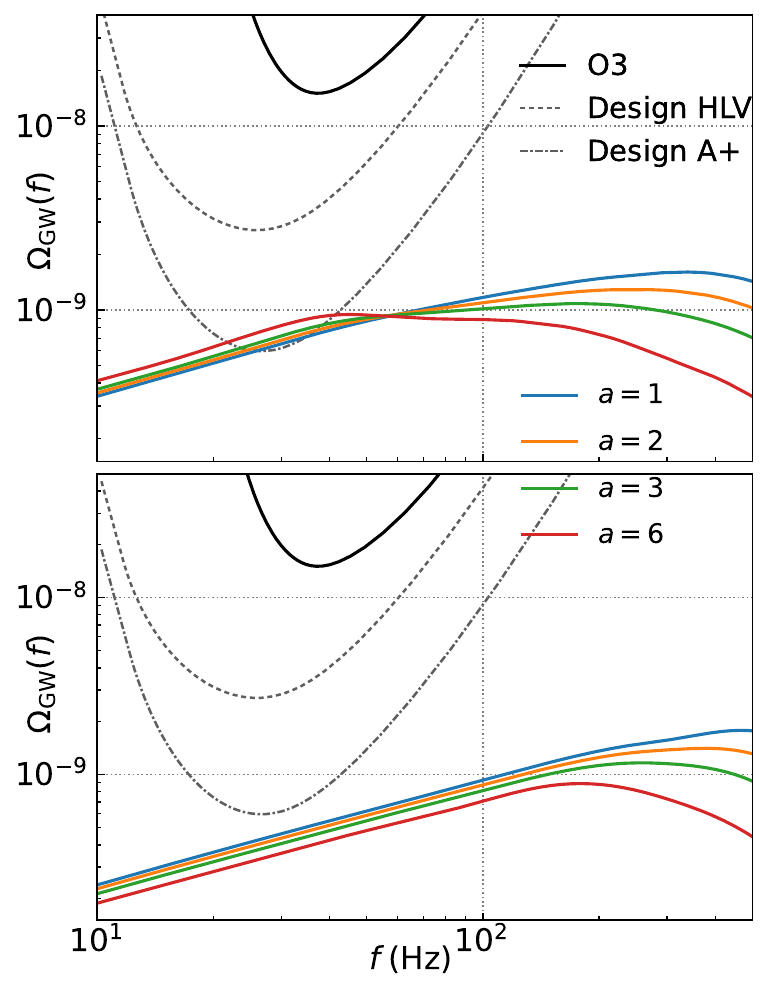}
\caption{Top: $\Omega_{\mathrm{GW}}(f)$ from the \textsc{cmc} model is shown for different values of the $a$ parameter after fixing $b=5$, $z_{\mathrm{peak}} = 5$, $f_{\mathrm{ev}}=80$, $\mu_r =2~\mathrm{pc}$, and $\sigma_r = 1.5\;\mathrm{pc}$. Black lines give $2\sigma$ power-law integrated (PI) curves for O3, projections for the HLV network at design sensitivity, and A+ detectors. Bottom: Same as top panel, but using the PLPP mass distribution instead of the BBH masses given by the \textsc{cmc} catalog.}
\label{fig:fig18}
\end{figure}

The parameter $a$ is associated with the behaviour of both $\mathcal{R}_{\mathrm{GC}}(z)$ and $\mathcal{R}_{\mathrm{dyn}}(z)$ at low redshifts. To investigate its impact on the SGWB, we compute the spectrum under both the \textsc{cmc} and PLPP models, 
fixing $z_{\mathrm{peak}}=5$, $b=5$, $\mu_r = 2~\mathrm{pc}$, and $\sigma_r = 1.5\;\mathrm{pc}$ while allowing $a$ to vary. The comparison between the top and bottom plots in Figure~\ref{fig:fig18}, where the peak positions of the SGWB spectra shift to lower frequencies under the \textsc{cmc} model as compared to its PLPP counterpart, indicates that the parameter $a$ also plays a nontrivial role when considering variations in the mass components distribution. This observation highlights an interesting possibility of jointly constraining parameters $a$ and $z_{\mathrm{peak}}$ (and perhaps $b$ too), when studying the redshift-dependent mass components distributions with SGWB search results~\cite{Callister:2020arv,Atal:2022zux}. We leave this for future work.

\section{GW Energy Spectrum}
\label{sec:waveform}
To approximate the GW energy spectrum $\frac{\mathrm{d}E_{\mathrm{GW}}}{\mathrm{d} f_r}$ for a coalescing BBH in the source frame, we utilize the phenomenological templates models proposed in \cite{Ajith:2009bn}. These models serve to approximate the inspiral, merger, and ringdown waveforms in the frequency Fourier space domain $f_r$, expressed in terms of the source properties including the chirp mass $\mathcal{M}_c$ of the GW sources
\begin{equation}
\frac{\mathrm{d} E}{\mathrm{d} f_r}=\frac{(G \pi)^{2 / 3} \mathcal{M}_c^{5 / 3}}{3} \Gamma\left(f_r\right).
\end{equation}
The function $\Gamma(f_r)$ encodes the evolution of the waveform as a function of frequency in different phases of the coalescence as
\begin{equation}
\Gamma\left(f_r\right)=\left\{\begin{array}{l}
f_r^{-1 / 3} ~ f_r<f_{\text {merg }}, \\
\frac{f_r^{2 / 3}}{f_{\text {merg }}}  ~ f_{\text {merg }} \leqslant f_r<f_{\text {ring }}, \\
\frac{f_r^2}{f_{\text {merg }} f_{\text {ring }}^{4 / 3}}\frac{1}{\left(1+\left(\frac{f_r-f_{\text {ring }}}{f_w / 2}\right)^2\right)^2} ~ f_{\text {ring }} \leqslant f_r<f_{\text {cut }},
\end{array}\right.
\end{equation}
where the merger $f_{\text{merg}}$, ringdown $f_{\text{ring}}$, cut $f_{\text{cut}}$ frequencies and the width of the Lorentzian function $f_w$ are given in given in Sec. IV of~\cite{Ajith:2009bn}.

\nocite{*}

\bibliography{ref}

\providecommand{\noopsort}[1]{}\providecommand{\singleletter}[1]{#1}%
\begin{thebibliography}{102}%
\makeatletter
\providecommand \@ifxundefined [1]{%
 \@ifx{#1\undefined}
}%
\providecommand \@ifnum [1]{%
 \ifnum #1\expandafter \@firstoftwo
 \else \expandafter \@secondoftwo
 \fi
}%
\providecommand \@ifx [1]{%
 \ifx #1\expandafter \@firstoftwo
 \else \expandafter \@secondoftwo
 \fi
}%
\providecommand \natexlab [1]{#1}%
\providecommand \enquote  [1]{``#1''}%
\providecommand \bibnamefont  [1]{#1}%
\providecommand \bibfnamefont [1]{#1}%
\providecommand \citenamefont [1]{#1}%
\providecommand \href@noop [0]{\@secondoftwo}%
\providecommand \href [0]{\begingroup \@sanitize@url \@href}%
\providecommand \@href[1]{\@@startlink{#1}\@@href}%
\providecommand \@@href[1]{\endgroup#1\@@endlink}%
\providecommand \@sanitize@url [0]{\catcode `\\12\catcode `\$12\catcode `\&12\catcode `\#12\catcode `\^12\catcode `\_12\catcode `\%12\relax}%
\providecommand \@@startlink[1]{}%
\providecommand \@@endlink[0]{}%
\providecommand \url  [0]{\begingroup\@sanitize@url \@url }%
\providecommand \@url [1]{\endgroup\@href {#1}{\urlprefix }}%
\providecommand \urlprefix  [0]{URL }%
\providecommand \Eprint [0]{\href }%
\providecommand \doibase [0]{http://dx.doi.org/}%
\providecommand \selectlanguage [0]{\@gobble}%
\providecommand \bibinfo  [0]{\@secondoftwo}%
\providecommand \bibfield  [0]{\@secondoftwo}%
\providecommand \translation [1]{[#1]}%
\providecommand \BibitemOpen [0]{}%
\providecommand \bibitemStop [0]{}%
\providecommand \bibitemNoStop [0]{.\EOS\space}%
\providecommand \EOS [0]{\spacefactor3000\relax}%
\providecommand \BibitemShut  [1]{\csname bibitem#1\endcsname}%
\let\auto@bib@innerbib\@empty
\bibitem [{\citenamefont {Aasi}\ \emph {et~al.}(2015)\citenamefont {Aasi} \emph {et~al.}}]{LIGOScientific:2014pky}%
  \BibitemOpen
  \bibfield  {author} {\bibinfo {author} {\bibfnamefont {J.}~\bibnamefont {Aasi}} \emph {et~al.} (\bibinfo {collaboration} {LIGO Scientific}),\ }\href {\doibase 10.1088/0264-9381/32/7/074001} {\bibfield  {journal} {\bibinfo  {journal} {Class. Quant. Grav.}\ }\textbf {\bibinfo {volume} {32}},\ \bibinfo {pages} {074001} (\bibinfo {year} {2015})},\ \Eprint {http://arxiv.org/abs/1411.4547} {arXiv:1411.4547 [gr-qc]} \BibitemShut {NoStop}%
\bibitem [{\citenamefont {Acernese}\ \emph {et~al.}(2015)\citenamefont {Acernese} \emph {et~al.}}]{VIRGO:2014yos}%
  \BibitemOpen
  \bibfield  {author} {\bibinfo {author} {\bibfnamefont {F.}~\bibnamefont {Acernese}} \emph {et~al.} (\bibinfo {collaboration} {VIRGO}),\ }\href {\doibase 10.1088/0264-9381/32/2/024001} {\bibfield  {journal} {\bibinfo  {journal} {Class. Quant. Grav.}\ }\textbf {\bibinfo {volume} {32}},\ \bibinfo {pages} {024001} (\bibinfo {year} {2015})},\ \Eprint {http://arxiv.org/abs/1408.3978} {arXiv:1408.3978 [gr-qc]} \BibitemShut {NoStop}%
\bibitem [{\citenamefont {Akutsu}\ \emph {et~al.}(2021)\citenamefont {Akutsu} \emph {et~al.}}]{KAGRA:2020tym}%
  \BibitemOpen
  \bibfield  {author} {\bibinfo {author} {\bibfnamefont {T.}~\bibnamefont {Akutsu}} \emph {et~al.} (\bibinfo {collaboration} {KAGRA}),\ }\href {\doibase 10.1093/ptep/ptaa125} {\bibfield  {journal} {\bibinfo  {journal} {PTEP}\ }\textbf {\bibinfo {volume} {2021}},\ \bibinfo {pages} {05A101} (\bibinfo {year} {2021})},\ \Eprint {http://arxiv.org/abs/2005.05574} {arXiv:2005.05574 [physics.ins-det]} \BibitemShut {NoStop}%
\bibitem [{\citenamefont {Abbott}\ \emph {et~al.}(2021{\natexlab{a}})\citenamefont {Abbott} \emph {et~al.}}]{LIGOScientific:2020ibl}%
  \BibitemOpen
  \bibfield  {author} {\bibinfo {author} {\bibfnamefont {R.}~\bibnamefont {Abbott}} \emph {et~al.} (\bibinfo {collaboration} {LIGO Scientific, Virgo}),\ }\href {\doibase 10.1103/PhysRevX.11.021053} {\bibfield  {journal} {\bibinfo  {journal} {Phys. Rev. X}\ }\textbf {\bibinfo {volume} {11}},\ \bibinfo {pages} {021053} (\bibinfo {year} {2021}{\natexlab{a}})},\ \Eprint {http://arxiv.org/abs/2010.14527} {arXiv:2010.14527 [gr-qc]} \BibitemShut {NoStop}%
\bibitem [{\citenamefont {Abbott}\ \emph {et~al.}(2021{\natexlab{b}})\citenamefont {Abbott} \emph {et~al.}}]{LIGOScientific:2021djp}%
  \BibitemOpen
  \bibfield  {author} {\bibinfo {author} {\bibfnamefont {R.}~\bibnamefont {Abbott}} \emph {et~al.} (\bibinfo {collaboration} {LIGO Scientific, VIRGO, KAGRA}),\ }\href@noop {} {\  (\bibinfo {year} {2021}{\natexlab{b}})},\ \Eprint {http://arxiv.org/abs/2111.03606} {arXiv:2111.03606 [gr-qc]} \BibitemShut {NoStop}%
\bibitem [{\citenamefont {Abbott}\ \emph {et~al.}(2021{\natexlab{c}})\citenamefont {Abbott} \emph {et~al.}}]{LIGOScientific:2021usb}%
  \BibitemOpen
  \bibfield  {author} {\bibinfo {author} {\bibfnamefont {R.}~\bibnamefont {Abbott}} \emph {et~al.} (\bibinfo {collaboration} {LIGO Scientific, VIRGO}),\ }\href@noop {} {\  (\bibinfo {year} {2021}{\natexlab{c}})},\ \Eprint {http://arxiv.org/abs/2108.01045} {arXiv:2108.01045 [gr-qc]} \BibitemShut {NoStop}%
\bibitem [{\citenamefont {Abbott}\ \emph {et~al.}(2021{\natexlab{d}})\citenamefont {Abbott} \emph {et~al.}}]{KAGRA:2021kbb}%
  \BibitemOpen
  \bibfield  {author} {\bibinfo {author} {\bibfnamefont {R.}~\bibnamefont {Abbott}} \emph {et~al.} (\bibinfo {collaboration} {KAGRA, Virgo, LIGO Scientific}),\ }\href {\doibase 10.1103/PhysRevD.104.022004} {\bibfield  {journal} {\bibinfo  {journal} {Phys. Rev. D}\ }\textbf {\bibinfo {volume} {104}},\ \bibinfo {pages} {022004} (\bibinfo {year} {2021}{\natexlab{d}})},\ \Eprint {http://arxiv.org/abs/2101.12130} {arXiv:2101.12130 [gr-qc]} \BibitemShut {NoStop}%
\bibitem [{\citenamefont {Kimball}\ \emph {et~al.}(2021)\citenamefont {Kimball} \emph {et~al.}}]{Kimball:2020qyd}%
  \BibitemOpen
  \bibfield  {author} {\bibinfo {author} {\bibfnamefont {C.}~\bibnamefont {Kimball}} \emph {et~al.},\ }\href {\doibase 10.3847/2041-8213/ac0aef} {\bibfield  {journal} {\bibinfo  {journal} {Astrophys. J. Lett.}\ }\textbf {\bibinfo {volume} {915}},\ \bibinfo {pages} {L35} (\bibinfo {year} {2021})},\ \Eprint {http://arxiv.org/abs/2011.05332} {arXiv:2011.05332 [astro-ph.HE]} \BibitemShut {NoStop}%
\bibitem [{\citenamefont {Zevin}\ \emph {et~al.}(2021)\citenamefont {Zevin}, \citenamefont {Bavera}, \citenamefont {Berry}, \citenamefont {Kalogera}, \citenamefont {Fragos}, \citenamefont {Marchant}, \citenamefont {Rodriguez}, \citenamefont {Antonini}, \citenamefont {Holz},\ and\ \citenamefont {Pankow}}]{Zevin:2020gbd}%
  \BibitemOpen
  \bibfield  {author} {\bibinfo {author} {\bibfnamefont {M.}~\bibnamefont {Zevin}}, \bibinfo {author} {\bibfnamefont {S.~S.}\ \bibnamefont {Bavera}}, \bibinfo {author} {\bibfnamefont {C.~P.~L.}\ \bibnamefont {Berry}}, \bibinfo {author} {\bibfnamefont {V.}~\bibnamefont {Kalogera}}, \bibinfo {author} {\bibfnamefont {T.}~\bibnamefont {Fragos}}, \bibinfo {author} {\bibfnamefont {P.}~\bibnamefont {Marchant}}, \bibinfo {author} {\bibfnamefont {C.~L.}\ \bibnamefont {Rodriguez}}, \bibinfo {author} {\bibfnamefont {F.}~\bibnamefont {Antonini}}, \bibinfo {author} {\bibfnamefont {D.~E.}\ \bibnamefont {Holz}}, \ and\ \bibinfo {author} {\bibfnamefont {C.}~\bibnamefont {Pankow}},\ }\href {\doibase 10.3847/1538-4357/abe40e} {\bibfield  {journal} {\bibinfo  {journal} {Astrophys. J.}\ }\textbf {\bibinfo {volume} {910}},\ \bibinfo {pages} {152} (\bibinfo {year} {2021})},\ \Eprint {http://arxiv.org/abs/2011.10057} {arXiv:2011.10057 [astro-ph.HE]} \BibitemShut {NoStop}%
\bibitem [{\citenamefont {Mapelli}\ \emph {et~al.}(2019)\citenamefont {Mapelli}, \citenamefont {Giacobbo}, \citenamefont {Santoliquido},\ and\ \citenamefont {Artale}}]{Mapelli:2019bnp}%
  \BibitemOpen
  \bibfield  {author} {\bibinfo {author} {\bibfnamefont {M.}~\bibnamefont {Mapelli}}, \bibinfo {author} {\bibfnamefont {N.}~\bibnamefont {Giacobbo}}, \bibinfo {author} {\bibfnamefont {F.}~\bibnamefont {Santoliquido}}, \ and\ \bibinfo {author} {\bibfnamefont {M.~C.}\ \bibnamefont {Artale}},\ }\href {\doibase 10.1093/mnras/stz1150} {\bibfield  {journal} {\bibinfo  {journal} {Mon. Not. Roy. Astron. Soc.}\ }\textbf {\bibinfo {volume} {487}},\ \bibinfo {pages} {2} (\bibinfo {year} {2019})},\ \Eprint {http://arxiv.org/abs/1902.01419} {arXiv:1902.01419 [astro-ph.HE]} \BibitemShut {NoStop}%
\bibitem [{\citenamefont {Baibhav}\ \emph {et~al.}(2019)\citenamefont {Baibhav}, \citenamefont {Berti}, \citenamefont {Gerosa}, \citenamefont {Mapelli}, \citenamefont {Giacobbo}, \citenamefont {Bouffanais},\ and\ \citenamefont {Di~Carlo}}]{Baibhav:2019gxm}%
  \BibitemOpen
  \bibfield  {author} {\bibinfo {author} {\bibfnamefont {V.}~\bibnamefont {Baibhav}}, \bibinfo {author} {\bibfnamefont {E.}~\bibnamefont {Berti}}, \bibinfo {author} {\bibfnamefont {D.}~\bibnamefont {Gerosa}}, \bibinfo {author} {\bibfnamefont {M.}~\bibnamefont {Mapelli}}, \bibinfo {author} {\bibfnamefont {N.}~\bibnamefont {Giacobbo}}, \bibinfo {author} {\bibfnamefont {Y.}~\bibnamefont {Bouffanais}}, \ and\ \bibinfo {author} {\bibfnamefont {U.~N.}\ \bibnamefont {Di~Carlo}},\ }\href {\doibase 10.1103/PhysRevD.100.064060} {\bibfield  {journal} {\bibinfo  {journal} {Phys. Rev. D}\ }\textbf {\bibinfo {volume} {100}},\ \bibinfo {pages} {064060} (\bibinfo {year} {2019})},\ \Eprint {http://arxiv.org/abs/1906.04197} {arXiv:1906.04197 [gr-qc]} \BibitemShut {NoStop}%
\bibitem [{\citenamefont {Broekgaarden}\ \emph {et~al.}(2021)\citenamefont {Broekgaarden} \emph {et~al.}}]{Broekgaarden:2021efa}%
  \BibitemOpen
  \bibfield  {author} {\bibinfo {author} {\bibfnamefont {F.~S.}\ \bibnamefont {Broekgaarden}} \emph {et~al.},\ }\href {\doibase 10.1093/mnras/stac1677} {\  (\bibinfo {year} {2021}),\ 10.1093/mnras/stac1677},\ \Eprint {http://arxiv.org/abs/2112.05763} {arXiv:2112.05763 [astro-ph.HE]} \BibitemShut {NoStop}%
\bibitem [{\citenamefont {van Son}\ \emph {et~al.}(2022)\citenamefont {van Son}, \citenamefont {de~Mink}, \citenamefont {Callister}, \citenamefont {Justham}, \citenamefont {Renzo}, \citenamefont {Wagg}, \citenamefont {Broekgaarden}, \citenamefont {Kummer}, \citenamefont {Pakmor},\ and\ \citenamefont {Mandel}}]{vanSon:2021zpk}%
  \BibitemOpen
  \bibfield  {author} {\bibinfo {author} {\bibfnamefont {L.~A.~C.}\ \bibnamefont {van Son}}, \bibinfo {author} {\bibfnamefont {S.~E.}\ \bibnamefont {de~Mink}}, \bibinfo {author} {\bibfnamefont {T.}~\bibnamefont {Callister}}, \bibinfo {author} {\bibfnamefont {S.}~\bibnamefont {Justham}}, \bibinfo {author} {\bibfnamefont {M.}~\bibnamefont {Renzo}}, \bibinfo {author} {\bibfnamefont {T.}~\bibnamefont {Wagg}}, \bibinfo {author} {\bibfnamefont {F.~S.}\ \bibnamefont {Broekgaarden}}, \bibinfo {author} {\bibfnamefont {F.}~\bibnamefont {Kummer}}, \bibinfo {author} {\bibfnamefont {R.}~\bibnamefont {Pakmor}}, \ and\ \bibinfo {author} {\bibfnamefont {I.}~\bibnamefont {Mandel}},\ }\href {\doibase 10.3847/1538-4357/ac64a3} {\bibfield  {journal} {\bibinfo  {journal} {Astrophys. J.}\ }\textbf {\bibinfo {volume} {931}},\ \bibinfo {pages} {17} (\bibinfo {year} {2022})},\ \Eprint {http://arxiv.org/abs/2110.01634} {arXiv:2110.01634 [astro-ph.HE]} \BibitemShut {NoStop}%
\bibitem [{\citenamefont {Mapelli}(2021)}]{Mapelli:2021taw}%
  \BibitemOpen
  \bibfield  {author} {\bibinfo {author} {\bibfnamefont {M.}~\bibnamefont {Mapelli}},\ }\enquote {\bibinfo {title} {{Formation Channels of Single and Binary Stellar-Mass Black Holes}},}\ \ (\bibinfo {year} {2021})\ \Eprint {http://arxiv.org/abs/2106.00699} {arXiv:2106.00699 [astro-ph.HE]} \BibitemShut {NoStop}%
\bibitem [{\citenamefont {Mandel}\ and\ \citenamefont {Broekgaarden}(2022)}]{Mandel:2021smh}%
  \BibitemOpen
  \bibfield  {author} {\bibinfo {author} {\bibfnamefont {I.}~\bibnamefont {Mandel}}\ and\ \bibinfo {author} {\bibfnamefont {F.~S.}\ \bibnamefont {Broekgaarden}},\ }\href {\doibase 10.1007/s41114-021-00034-3} {\bibfield  {journal} {\bibinfo  {journal} {Living Rev. Rel.}\ }\textbf {\bibinfo {volume} {25}},\ \bibinfo {pages} {1} (\bibinfo {year} {2022})},\ \Eprint {http://arxiv.org/abs/2107.14239} {arXiv:2107.14239 [astro-ph.HE]} \BibitemShut {NoStop}%
\bibitem [{\citenamefont {Belczynski}\ \emph {et~al.}(2001)\citenamefont {Belczynski}, \citenamefont {Kalogera},\ and\ \citenamefont {Bulik}}]{Belczynski:2001uc}%
  \BibitemOpen
  \bibfield  {author} {\bibinfo {author} {\bibfnamefont {K.}~\bibnamefont {Belczynski}}, \bibinfo {author} {\bibfnamefont {V.}~\bibnamefont {Kalogera}}, \ and\ \bibinfo {author} {\bibfnamefont {T.}~\bibnamefont {Bulik}},\ }\href {\doibase 10.1086/340304} {\bibfield  {journal} {\bibinfo  {journal} {Astrophys. J.}\ }\textbf {\bibinfo {volume} {572}},\ \bibinfo {pages} {407} (\bibinfo {year} {2001})},\ \Eprint {http://arxiv.org/abs/astro-ph/0111452} {arXiv:astro-ph/0111452} \BibitemShut {NoStop}%
\bibitem [{\citenamefont {Dominik}\ \emph {et~al.}(2012)\citenamefont {Dominik}, \citenamefont {Belczynski}, \citenamefont {Fryer}, \citenamefont {Holz}, \citenamefont {Berti}, \citenamefont {Bulik}, \citenamefont {Mandel},\ and\ \citenamefont {O'Shaughnessy}}]{Dominik:2012kk}%
  \BibitemOpen
  \bibfield  {author} {\bibinfo {author} {\bibfnamefont {M.}~\bibnamefont {Dominik}}, \bibinfo {author} {\bibfnamefont {K.}~\bibnamefont {Belczynski}}, \bibinfo {author} {\bibfnamefont {C.}~\bibnamefont {Fryer}}, \bibinfo {author} {\bibfnamefont {D.}~\bibnamefont {Holz}}, \bibinfo {author} {\bibfnamefont {E.}~\bibnamefont {Berti}}, \bibinfo {author} {\bibfnamefont {T.}~\bibnamefont {Bulik}}, \bibinfo {author} {\bibfnamefont {I.}~\bibnamefont {Mandel}}, \ and\ \bibinfo {author} {\bibfnamefont {R.}~\bibnamefont {O'Shaughnessy}},\ }\href {\doibase 10.1088/0004-637X/759/1/52} {\bibfield  {journal} {\bibinfo  {journal} {Astrophys. J.}\ }\textbf {\bibinfo {volume} {759}},\ \bibinfo {pages} {52} (\bibinfo {year} {2012})},\ \Eprint {http://arxiv.org/abs/1202.4901} {arXiv:1202.4901 [astro-ph.HE]} \BibitemShut {NoStop}%
\bibitem [{\citenamefont {Belczynski}\ \emph {et~al.}(2020)\citenamefont {Belczynski} \emph {et~al.}}]{Belczynski:2017gds}%
  \BibitemOpen
  \bibfield  {author} {\bibinfo {author} {\bibfnamefont {K.}~\bibnamefont {Belczynski}} \emph {et~al.},\ }\href {\doibase 10.1051/0004-6361/201936528} {\bibfield  {journal} {\bibinfo  {journal} {Astron. Astrophys.}\ }\textbf {\bibinfo {volume} {636}},\ \bibinfo {pages} {A104} (\bibinfo {year} {2020})},\ \Eprint {http://arxiv.org/abs/1706.07053} {arXiv:1706.07053 [astro-ph.HE]} \BibitemShut {NoStop}%
\bibitem [{\citenamefont {Rodriguez}\ \emph {et~al.}(2016)\citenamefont {Rodriguez}, \citenamefont {Chatterjee},\ and\ \citenamefont {Rasio}}]{Rodriguez:2016kxx}%
  \BibitemOpen
  \bibfield  {author} {\bibinfo {author} {\bibfnamefont {C.~L.}\ \bibnamefont {Rodriguez}}, \bibinfo {author} {\bibfnamefont {S.}~\bibnamefont {Chatterjee}}, \ and\ \bibinfo {author} {\bibfnamefont {F.~A.}\ \bibnamefont {Rasio}},\ }\href {\doibase 10.1103/PhysRevD.93.084029} {\bibfield  {journal} {\bibinfo  {journal} {Phys. Rev. D}\ }\textbf {\bibinfo {volume} {93}},\ \bibinfo {pages} {084029} (\bibinfo {year} {2016})},\ \Eprint {http://arxiv.org/abs/1602.02444} {arXiv:1602.02444 [astro-ph.HE]} \BibitemShut {NoStop}%
\bibitem [{\citenamefont {Rodriguez}\ \emph {et~al.}(2018)\citenamefont {Rodriguez}, \citenamefont {Amaro-Seoane}, \citenamefont {Chatterjee},\ and\ \citenamefont {Rasio}}]{Rodriguez:2017pec}%
  \BibitemOpen
  \bibfield  {author} {\bibinfo {author} {\bibfnamefont {C.~L.}\ \bibnamefont {Rodriguez}}, \bibinfo {author} {\bibfnamefont {P.}~\bibnamefont {Amaro-Seoane}}, \bibinfo {author} {\bibfnamefont {S.}~\bibnamefont {Chatterjee}}, \ and\ \bibinfo {author} {\bibfnamefont {F.~A.}\ \bibnamefont {Rasio}},\ }\href {\doibase 10.1103/PhysRevLett.120.151101} {\bibfield  {journal} {\bibinfo  {journal} {Phys. Rev. Lett.}\ }\textbf {\bibinfo {volume} {120}},\ \bibinfo {pages} {151101} (\bibinfo {year} {2018})},\ \Eprint {http://arxiv.org/abs/1712.04937} {arXiv:1712.04937 [astro-ph.HE]} \BibitemShut {NoStop}%
\bibitem [{\citenamefont {Di~Carlo}\ \emph {et~al.}(2019)\citenamefont {Di~Carlo}, \citenamefont {Giacobbo}, \citenamefont {Mapelli}, \citenamefont {Pasquato}, \citenamefont {Spera}, \citenamefont {Wang},\ and\ \citenamefont {Haardt}}]{DiCarlo:2019pmf}%
  \BibitemOpen
  \bibfield  {author} {\bibinfo {author} {\bibfnamefont {U.~N.}\ \bibnamefont {Di~Carlo}}, \bibinfo {author} {\bibfnamefont {N.}~\bibnamefont {Giacobbo}}, \bibinfo {author} {\bibfnamefont {M.}~\bibnamefont {Mapelli}}, \bibinfo {author} {\bibfnamefont {M.}~\bibnamefont {Pasquato}}, \bibinfo {author} {\bibfnamefont {M.}~\bibnamefont {Spera}}, \bibinfo {author} {\bibfnamefont {L.}~\bibnamefont {Wang}}, \ and\ \bibinfo {author} {\bibfnamefont {F.}~\bibnamefont {Haardt}},\ }\href {\doibase 10.1093/mnras/stz1453} {\bibfield  {journal} {\bibinfo  {journal} {Mon. Not. Roy. Astron. Soc.}\ }\textbf {\bibinfo {volume} {487}},\ \bibinfo {pages} {2947} (\bibinfo {year} {2019})},\ \Eprint {http://arxiv.org/abs/1901.00863} {arXiv:1901.00863 [astro-ph.HE]} \BibitemShut {NoStop}%
\bibitem [{\citenamefont {Antonini}\ \emph {et~al.}(2019)\citenamefont {Antonini}, \citenamefont {Gieles},\ and\ \citenamefont {Gualandris}}]{Antonini:2018auk}%
  \BibitemOpen
  \bibfield  {author} {\bibinfo {author} {\bibfnamefont {F.}~\bibnamefont {Antonini}}, \bibinfo {author} {\bibfnamefont {M.}~\bibnamefont {Gieles}}, \ and\ \bibinfo {author} {\bibfnamefont {A.}~\bibnamefont {Gualandris}},\ }\href {\doibase 10.1093/mnras/stz1149} {\bibfield  {journal} {\bibinfo  {journal} {Mon. Not. Roy. Astron. Soc.}\ }\textbf {\bibinfo {volume} {486}},\ \bibinfo {pages} {5008} (\bibinfo {year} {2019})},\ \Eprint {http://arxiv.org/abs/1811.03640} {arXiv:1811.03640 [astro-ph.HE]} \BibitemShut {NoStop}%
\bibitem [{\citenamefont {Mapelli}\ \emph {et~al.}(2022)\citenamefont {Mapelli}, \citenamefont {Bouffanais}, \citenamefont {Santoliquido}, \citenamefont {Sedda},\ and\ \citenamefont {Artale}}]{Mapelli:2021gyv}%
  \BibitemOpen
  \bibfield  {author} {\bibinfo {author} {\bibfnamefont {M.}~\bibnamefont {Mapelli}}, \bibinfo {author} {\bibfnamefont {Y.}~\bibnamefont {Bouffanais}}, \bibinfo {author} {\bibfnamefont {F.}~\bibnamefont {Santoliquido}}, \bibinfo {author} {\bibfnamefont {M.~A.}\ \bibnamefont {Sedda}}, \ and\ \bibinfo {author} {\bibfnamefont {M.~C.}\ \bibnamefont {Artale}},\ }\href {\doibase 10.1093/mnras/stac422} {\bibfield  {journal} {\bibinfo  {journal} {Mon. Not. Roy. Astron. Soc.}\ }\textbf {\bibinfo {volume} {511}},\ \bibinfo {pages} {5797} (\bibinfo {year} {2022})},\ \Eprint {http://arxiv.org/abs/2109.06222} {arXiv:2109.06222 [astro-ph.HE]} \BibitemShut {NoStop}%
\bibitem [{\citenamefont {{Fragione}}\ \emph {et~al.}(2022)\citenamefont {{Fragione}}, \citenamefont {{Kocsis}}, \citenamefont {{Rasio}},\ and\ \citenamefont {{Silk}}}]{FragioneKocsisetal2022}%
  \BibitemOpen
  \bibfield  {author} {\bibinfo {author} {\bibfnamefont {G.}~\bibnamefont {{Fragione}}}, \bibinfo {author} {\bibfnamefont {B.}~\bibnamefont {{Kocsis}}}, \bibinfo {author} {\bibfnamefont {F.~A.}\ \bibnamefont {{Rasio}}}, \ and\ \bibinfo {author} {\bibfnamefont {J.}~\bibnamefont {{Silk}}},\ }\href {\doibase 10.3847/1538-4357/ac5026} {\bibfield  {journal} {\bibinfo  {journal} {\apj}\ }\textbf {\bibinfo {volume} {927}},\ \bibinfo {eid} {231} (\bibinfo {year} {2022})},\ \Eprint {http://arxiv.org/abs/2107.04639} {arXiv:2107.04639 [astro-ph.GA]} \BibitemShut {NoStop}%
\bibitem [{\citenamefont {{Hoang}}\ \emph {et~al.}(2018)\citenamefont {{Hoang}}, \citenamefont {{Naoz}}, \citenamefont {{Kocsis}}, \citenamefont {{Rasio}},\ and\ \citenamefont {et~al.}}]{HoangNaoz2018}%
  \BibitemOpen
  \bibfield  {author} {\bibinfo {author} {\bibfnamefont {B.-M.}\ \bibnamefont {{Hoang}}}, \bibinfo {author} {\bibfnamefont {S.}~\bibnamefont {{Naoz}}}, \bibinfo {author} {\bibfnamefont {B.}~\bibnamefont {{Kocsis}}}, \bibinfo {author} {\bibfnamefont {F.~A.}\ \bibnamefont {{Rasio}}}, \ and\ \bibinfo {author} {\bibnamefont {et~al.}},\ }\href {\doibase 10.3847/1538-4357/aaafce} {\bibfield  {journal} {\bibinfo  {journal} {\apj}\ }\textbf {\bibinfo {volume} {856}},\ \bibinfo {eid} {140} (\bibinfo {year} {2018})},\ \Eprint {http://arxiv.org/abs/1706.09896} {arXiv:1706.09896 [astro-ph.HE]} \BibitemShut {NoStop}%
\bibitem [{\citenamefont {{Fragione}}\ \emph {et~al.}(2020)\citenamefont {{Fragione}}, \citenamefont {{Loeb}},\ and\ \citenamefont {{Rasio}}}]{FragioneLoebRasio2020}%
  \BibitemOpen
  \bibfield  {author} {\bibinfo {author} {\bibfnamefont {G.}~\bibnamefont {{Fragione}}}, \bibinfo {author} {\bibfnamefont {A.}~\bibnamefont {{Loeb}}}, \ and\ \bibinfo {author} {\bibfnamefont {F.~A.}\ \bibnamefont {{Rasio}}},\ }\href {\doibase 10.3847/2041-8213/ab9093} {\bibfield  {journal} {\bibinfo  {journal} {\apjl}\ }\textbf {\bibinfo {volume} {895}},\ \bibinfo {eid} {L15} (\bibinfo {year} {2020})},\ \Eprint {http://arxiv.org/abs/2002.11278} {arXiv:2002.11278 [astro-ph.GA]} \BibitemShut {NoStop}%
\bibitem [{\citenamefont {{Bartos}}\ \emph {et~al.}(2017)\citenamefont {{Bartos}}, \citenamefont {{Kocsis}}, \citenamefont {{Haiman}},\ and\ \citenamefont {{M{\'a}rka}}}]{BartosKocsis2017}%
  \BibitemOpen
  \bibfield  {author} {\bibinfo {author} {\bibfnamefont {I.}~\bibnamefont {{Bartos}}}, \bibinfo {author} {\bibfnamefont {B.}~\bibnamefont {{Kocsis}}}, \bibinfo {author} {\bibfnamefont {Z.}~\bibnamefont {{Haiman}}}, \ and\ \bibinfo {author} {\bibfnamefont {S.}~\bibnamefont {{M{\'a}rka}}},\ }\href {\doibase 10.3847/1538-4357/835/2/165} {\bibfield  {journal} {\bibinfo  {journal} {\apj}\ }\textbf {\bibinfo {volume} {835}},\ \bibinfo {eid} {165} (\bibinfo {year} {2017})},\ \Eprint {http://arxiv.org/abs/1602.03831} {arXiv:1602.03831 [astro-ph.HE]} \BibitemShut {NoStop}%
\bibitem [{\citenamefont {Sasaki}\ \emph {et~al.}(2016)\citenamefont {Sasaki}, \citenamefont {Suyama}, \citenamefont {Tanaka},\ and\ \citenamefont {Yokoyama}}]{Sasaki:2016jop}%
  \BibitemOpen
  \bibfield  {author} {\bibinfo {author} {\bibfnamefont {M.}~\bibnamefont {Sasaki}}, \bibinfo {author} {\bibfnamefont {T.}~\bibnamefont {Suyama}}, \bibinfo {author} {\bibfnamefont {T.}~\bibnamefont {Tanaka}}, \ and\ \bibinfo {author} {\bibfnamefont {S.}~\bibnamefont {Yokoyama}},\ }\href {\doibase 10.1103/PhysRevLett.117.061101} {\bibfield  {journal} {\bibinfo  {journal} {Phys. Rev. Lett.}\ }\textbf {\bibinfo {volume} {117}},\ \bibinfo {pages} {061101} (\bibinfo {year} {2016})},\ \bibinfo {note} {[Erratum: Phys.Rev.Lett. 121, 059901 (2018)]},\ \Eprint {http://arxiv.org/abs/1603.08338} {arXiv:1603.08338 [astro-ph.CO]} \BibitemShut {NoStop}%
\bibitem [{\citenamefont {Ali-Ha\"\i{}moud}\ and\ \citenamefont {Kamionkowski}(2017)}]{Ali-Haimoud:2016mbv}%
  \BibitemOpen
  \bibfield  {author} {\bibinfo {author} {\bibfnamefont {Y.}~\bibnamefont {Ali-Ha\"\i{}moud}}\ and\ \bibinfo {author} {\bibfnamefont {M.}~\bibnamefont {Kamionkowski}},\ }\href {\doibase 10.1103/PhysRevD.95.043534} {\bibfield  {journal} {\bibinfo  {journal} {Phys. Rev. D}\ }\textbf {\bibinfo {volume} {95}},\ \bibinfo {pages} {043534} (\bibinfo {year} {2017})},\ \Eprint {http://arxiv.org/abs/1612.05644} {arXiv:1612.05644 [astro-ph.CO]} \BibitemShut {NoStop}%
\bibitem [{\citenamefont {Bird}\ \emph {et~al.}(2016)\citenamefont {Bird}, \citenamefont {Cholis}, \citenamefont {Mu\~noz}, \citenamefont {Ali-Ha\"\i{}moud}, \citenamefont {Kamionkowski}, \citenamefont {Kovetz}, \citenamefont {Raccanelli},\ and\ \citenamefont {Riess}}]{Bird:2016dcv}%
  \BibitemOpen
  \bibfield  {author} {\bibinfo {author} {\bibfnamefont {S.}~\bibnamefont {Bird}}, \bibinfo {author} {\bibfnamefont {I.}~\bibnamefont {Cholis}}, \bibinfo {author} {\bibfnamefont {J.~B.}\ \bibnamefont {Mu\~noz}}, \bibinfo {author} {\bibfnamefont {Y.}~\bibnamefont {Ali-Ha\"\i{}moud}}, \bibinfo {author} {\bibfnamefont {M.}~\bibnamefont {Kamionkowski}}, \bibinfo {author} {\bibfnamefont {E.~D.}\ \bibnamefont {Kovetz}}, \bibinfo {author} {\bibfnamefont {A.}~\bibnamefont {Raccanelli}}, \ and\ \bibinfo {author} {\bibfnamefont {A.~G.}\ \bibnamefont {Riess}},\ }\href {\doibase 10.1103/PhysRevLett.116.201301} {\bibfield  {journal} {\bibinfo  {journal} {Phys. Rev. Lett.}\ }\textbf {\bibinfo {volume} {116}},\ \bibinfo {pages} {201301} (\bibinfo {year} {2016})},\ \Eprint {http://arxiv.org/abs/1603.00464} {arXiv:1603.00464 [astro-ph.CO]} \BibitemShut {NoStop}%
\bibitem [{\citenamefont {Clesse}\ and\ \citenamefont {Garc\'\i{}a-Bellido}(2017)}]{Clesse:2016vqa}%
  \BibitemOpen
  \bibfield  {author} {\bibinfo {author} {\bibfnamefont {S.}~\bibnamefont {Clesse}}\ and\ \bibinfo {author} {\bibfnamefont {J.}~\bibnamefont {Garc\'\i{}a-Bellido}},\ }\href {\doibase 10.1016/j.dark.2016.10.002} {\bibfield  {journal} {\bibinfo  {journal} {Phys. Dark Univ.}\ }\textbf {\bibinfo {volume} {15}},\ \bibinfo {pages} {142} (\bibinfo {year} {2017})},\ \Eprint {http://arxiv.org/abs/1603.05234} {arXiv:1603.05234 [astro-ph.CO]} \BibitemShut {NoStop}%
\bibitem [{\citenamefont {Abbott}\ \emph {et~al.}(2020{\natexlab{a}})\citenamefont {Abbott} \emph {et~al.}}]{LIGOScientific:2020iuh}%
  \BibitemOpen
  \bibfield  {author} {\bibinfo {author} {\bibfnamefont {R.}~\bibnamefont {Abbott}} \emph {et~al.} (\bibinfo {collaboration} {LIGO Scientific, Virgo}),\ }\href {\doibase 10.1103/PhysRevLett.125.101102} {\bibfield  {journal} {\bibinfo  {journal} {Phys. Rev. Lett.}\ }\textbf {\bibinfo {volume} {125}},\ \bibinfo {pages} {101102} (\bibinfo {year} {2020}{\natexlab{a}})},\ \Eprint {http://arxiv.org/abs/2009.01075} {arXiv:2009.01075 [gr-qc]} \BibitemShut {NoStop}%
\bibitem [{\citenamefont {Abbott}\ \emph {et~al.}(2020{\natexlab{b}})\citenamefont {Abbott} \emph {et~al.}}]{LIGOScientific:2020ufj}%
  \BibitemOpen
  \bibfield  {author} {\bibinfo {author} {\bibfnamefont {R.}~\bibnamefont {Abbott}} \emph {et~al.} (\bibinfo {collaboration} {LIGO Scientific, Virgo}),\ }\href {\doibase 10.3847/2041-8213/aba493} {\bibfield  {journal} {\bibinfo  {journal} {Astrophys. J. Lett.}\ }\textbf {\bibinfo {volume} {900}},\ \bibinfo {pages} {L13} (\bibinfo {year} {2020}{\natexlab{b}})},\ \Eprint {http://arxiv.org/abs/2009.01190} {arXiv:2009.01190 [astro-ph.HE]} \BibitemShut {NoStop}%
\bibitem [{\citenamefont {{Belczynski}}\ \emph {et~al.}(2016)\citenamefont {{Belczynski}}, \citenamefont {{Heger}}, \citenamefont {{Gladysz}}, \citenamefont {{Ruiter}},\ and\ \citenamefont {et~al.}}]{BelczynskiHeger2016}%
  \BibitemOpen
  \bibfield  {author} {\bibinfo {author} {\bibfnamefont {K.}~\bibnamefont {{Belczynski}}}, \bibinfo {author} {\bibfnamefont {A.}~\bibnamefont {{Heger}}}, \bibinfo {author} {\bibfnamefont {W.}~\bibnamefont {{Gladysz}}}, \bibinfo {author} {\bibfnamefont {A.~J.}\ \bibnamefont {{Ruiter}}}, \ and\ \bibinfo {author} {\bibnamefont {et~al.}},\ }\href {\doibase 10.1051/0004-6361/201628980} {\bibfield  {journal} {\bibinfo  {journal} {\aap}\ }\textbf {\bibinfo {volume} {594}},\ \bibinfo {eid} {A97} (\bibinfo {year} {2016})},\ \Eprint {http://arxiv.org/abs/1607.03116} {arXiv:1607.03116 [astro-ph.HE]} \BibitemShut {NoStop}%
\bibitem [{\citenamefont {Spera}\ and\ \citenamefont {Mapelli}(2017)}]{Spera:2017fyx}%
  \BibitemOpen
  \bibfield  {author} {\bibinfo {author} {\bibfnamefont {M.}~\bibnamefont {Spera}}\ and\ \bibinfo {author} {\bibfnamefont {M.}~\bibnamefont {Mapelli}},\ }\href {\doibase 10.1093/mnras/stx1576} {\bibfield  {journal} {\bibinfo  {journal} {Mon. Not. Roy. Astron. Soc.}\ }\textbf {\bibinfo {volume} {470}},\ \bibinfo {pages} {4739} (\bibinfo {year} {2017})},\ \Eprint {http://arxiv.org/abs/1706.06109} {arXiv:1706.06109 [astro-ph.SR]} \BibitemShut {NoStop}%
\bibitem [{\citenamefont {Stevenson}\ \emph {et~al.}(2019)\citenamefont {Stevenson}, \citenamefont {Sampson}, \citenamefont {Powell}, \citenamefont {Vigna-G\'omez}, \citenamefont {Neijssel}, \citenamefont {Sz\'ecsi},\ and\ \citenamefont {Mandel}}]{Stevenson:2019rcw}%
  \BibitemOpen
  \bibfield  {author} {\bibinfo {author} {\bibfnamefont {S.}~\bibnamefont {Stevenson}}, \bibinfo {author} {\bibfnamefont {M.}~\bibnamefont {Sampson}}, \bibinfo {author} {\bibfnamefont {J.}~\bibnamefont {Powell}}, \bibinfo {author} {\bibfnamefont {A.}~\bibnamefont {Vigna-G\'omez}}, \bibinfo {author} {\bibfnamefont {C.~J.}\ \bibnamefont {Neijssel}}, \bibinfo {author} {\bibfnamefont {D.}~\bibnamefont {Sz\'ecsi}}, \ and\ \bibinfo {author} {\bibfnamefont {I.}~\bibnamefont {Mandel}},\ }\href {\doibase 10.3847/1538-4357/ab3981} {\  (\bibinfo {year} {2019}),\ 10.3847/1538-4357/ab3981},\ \Eprint {http://arxiv.org/abs/1904.02821} {arXiv:1904.02821 [astro-ph.HE]} \BibitemShut {NoStop}%
\bibitem [{\citenamefont {Fishbach}\ and\ \citenamefont {Holz}(2020)}]{Fishbach:2020qag}%
  \BibitemOpen
  \bibfield  {author} {\bibinfo {author} {\bibfnamefont {M.}~\bibnamefont {Fishbach}}\ and\ \bibinfo {author} {\bibfnamefont {D.~E.}\ \bibnamefont {Holz}},\ }\href {\doibase 10.3847/2041-8213/abc827} {\bibfield  {journal} {\bibinfo  {journal} {Astrophys. J. Lett.}\ }\textbf {\bibinfo {volume} {904}},\ \bibinfo {pages} {L26} (\bibinfo {year} {2020})},\ \Eprint {http://arxiv.org/abs/2009.05472} {arXiv:2009.05472 [astro-ph.HE]} \BibitemShut {NoStop}%
\bibitem [{\citenamefont {Fragione}\ \emph {et~al.}(2020)\citenamefont {Fragione}, \citenamefont {Loeb},\ and\ \citenamefont {Rasio}}]{Fragione:2020han}%
  \BibitemOpen
  \bibfield  {author} {\bibinfo {author} {\bibfnamefont {G.}~\bibnamefont {Fragione}}, \bibinfo {author} {\bibfnamefont {A.}~\bibnamefont {Loeb}}, \ and\ \bibinfo {author} {\bibfnamefont {F.~A.}\ \bibnamefont {Rasio}},\ }\href {\doibase 10.3847/2041-8213/abbc0a} {\bibfield  {journal} {\bibinfo  {journal} {Astrophys. J. Lett.}\ }\textbf {\bibinfo {volume} {902}},\ \bibinfo {pages} {L26} (\bibinfo {year} {2020})},\ \Eprint {http://arxiv.org/abs/2009.05065} {arXiv:2009.05065 [astro-ph.GA]} \BibitemShut {NoStop}%
\bibitem [{\citenamefont {Liu}\ and\ \citenamefont {Lai}(2021)}]{Liu:2020gif}%
  \BibitemOpen
  \bibfield  {author} {\bibinfo {author} {\bibfnamefont {B.}~\bibnamefont {Liu}}\ and\ \bibinfo {author} {\bibfnamefont {D.}~\bibnamefont {Lai}},\ }\href {\doibase 10.1093/mnras/stab178} {\bibfield  {journal} {\bibinfo  {journal} {Mon. Not. Roy. Astron. Soc.}\ }\textbf {\bibinfo {volume} {502}},\ \bibinfo {pages} {2049} (\bibinfo {year} {2021})},\ \Eprint {http://arxiv.org/abs/2009.10068} {arXiv:2009.10068 [astro-ph.HE]} \BibitemShut {NoStop}%
\bibitem [{\citenamefont {{Fragione}}\ and\ \citenamefont {{Rasio}}(2023)}]{FragioneRasio2023}%
  \BibitemOpen
  \bibfield  {author} {\bibinfo {author} {\bibfnamefont {G.}~\bibnamefont {{Fragione}}}\ and\ \bibinfo {author} {\bibfnamefont {F.~A.}\ \bibnamefont {{Rasio}}},\ }\href {\doibase 10.3847/1538-4357/acd9c9} {\bibfield  {journal} {\bibinfo  {journal} {\apj}\ }\textbf {\bibinfo {volume} {951}},\ \bibinfo {eid} {129} (\bibinfo {year} {2023})},\ \Eprint {http://arxiv.org/abs/2302.11613} {arXiv:2302.11613 [astro-ph.GA]} \BibitemShut {NoStop}%
\bibitem [{\citenamefont {{Forbes}}\ \emph {et~al.}(2018)\citenamefont {{Forbes}}, \citenamefont {{Bastian}}, \citenamefont {{Gieles}}, \citenamefont {{Crain}}, \citenamefont {{Kruijssen}}, \citenamefont {{Larsen}}, \citenamefont {{Ploeckinger}}, \citenamefont {{Agertz}}, \citenamefont {{Trenti}}, \citenamefont {{Ferguson}},\ and\ \citenamefont {et~al.}}]{2018RSPSA.47470616F}%
  \BibitemOpen
  \bibfield  {author} {\bibinfo {author} {\bibfnamefont {D.~A.}\ \bibnamefont {{Forbes}}}, \bibinfo {author} {\bibfnamefont {N.}~\bibnamefont {{Bastian}}}, \bibinfo {author} {\bibfnamefont {M.}~\bibnamefont {{Gieles}}}, \bibinfo {author} {\bibfnamefont {R.~A.}\ \bibnamefont {{Crain}}}, \bibinfo {author} {\bibfnamefont {J.~M.~D.}\ \bibnamefont {{Kruijssen}}}, \bibinfo {author} {\bibfnamefont {S.~S.}\ \bibnamefont {{Larsen}}}, \bibinfo {author} {\bibfnamefont {S.}~\bibnamefont {{Ploeckinger}}}, \bibinfo {author} {\bibfnamefont {O.}~\bibnamefont {{Agertz}}}, \bibinfo {author} {\bibfnamefont {M.}~\bibnamefont {{Trenti}}}, \bibinfo {author} {\bibfnamefont {A.~M.~N.}\ \bibnamefont {{Ferguson}}}, \ and\ \bibinfo {author} {\bibnamefont {et~al.}},\ }\href {\doibase 10.1098/rspa.2017.061610.48550/arXiv.1801.05818} {\bibfield  {journal} {\bibinfo  {journal} {Proceedings of the Royal Society of London Series A}\ }\textbf {\bibinfo {volume} {474}},\ \bibinfo {eid} {20170616} (\bibinfo {year} {2018})},\ \Eprint
  {http://arxiv.org/abs/1801.05818} {arXiv:1801.05818 [astro-ph.GA]} \BibitemShut {NoStop}%
\bibitem [{\citenamefont {{Gratton}}\ \emph {et~al.}(2019)\citenamefont {{Gratton}}, \citenamefont {{Bragaglia}}, \citenamefont {{Carretta}}, \citenamefont {{D'Orazi}}, \citenamefont {{Lucatello}},\ and\ \citenamefont {{Sollima}}}]{2019A&ARv..27....8G}%
  \BibitemOpen
  \bibfield  {author} {\bibinfo {author} {\bibfnamefont {R.}~\bibnamefont {{Gratton}}}, \bibinfo {author} {\bibfnamefont {A.}~\bibnamefont {{Bragaglia}}}, \bibinfo {author} {\bibfnamefont {E.}~\bibnamefont {{Carretta}}}, \bibinfo {author} {\bibfnamefont {V.}~\bibnamefont {{D'Orazi}}}, \bibinfo {author} {\bibfnamefont {S.}~\bibnamefont {{Lucatello}}}, \ and\ \bibinfo {author} {\bibfnamefont {A.}~\bibnamefont {{Sollima}}},\ }\href {\doibase 10.1007/s00159-019-0119-3} {\bibfield  {journal} {\bibinfo  {journal} {\aapr}\ }\textbf {\bibinfo {volume} {27}},\ \bibinfo {eid} {8} (\bibinfo {year} {2019})},\ \Eprint {http://arxiv.org/abs/1911.02835} {arXiv:1911.02835 [astro-ph.SR]} \BibitemShut {NoStop}%
\bibitem [{\citenamefont {{Portegies Zwart}}\ \emph {et~al.}(2010)\citenamefont {{Portegies Zwart}}, \citenamefont {{McMillan}},\ and\ \citenamefont {{Gieles}}}]{PortegiesZwartMcMillan2010}%
  \BibitemOpen
  \bibfield  {author} {\bibinfo {author} {\bibfnamefont {S.~F.}\ \bibnamefont {{Portegies Zwart}}}, \bibinfo {author} {\bibfnamefont {S.~L.~W.}\ \bibnamefont {{McMillan}}}, \ and\ \bibinfo {author} {\bibfnamefont {M.}~\bibnamefont {{Gieles}}},\ }\href {\doibase 10.1146/annurev-astro-081309-130834} {\bibfield  {journal} {\bibinfo  {journal} {\araa}\ }\textbf {\bibinfo {volume} {48}},\ \bibinfo {pages} {431} (\bibinfo {year} {2010})},\ \Eprint {http://arxiv.org/abs/1002.1961} {arXiv:1002.1961 [astro-ph.GA]} \BibitemShut {NoStop}%
\bibitem [{\citenamefont {Fishbach}\ and\ \citenamefont {Fragione}(2023)}]{Fishbach:2023xws}%
  \BibitemOpen
  \bibfield  {author} {\bibinfo {author} {\bibfnamefont {M.}~\bibnamefont {Fishbach}}\ and\ \bibinfo {author} {\bibfnamefont {G.}~\bibnamefont {Fragione}},\ }\href {\doibase 10.1093/mnras/stad1364} {\bibfield  {journal} {\bibinfo  {journal} {Mon. Not. Roy. Astron. Soc.}\ }\textbf {\bibinfo {volume} {522}},\ \bibinfo {pages} {5546} (\bibinfo {year} {2023})},\ \Eprint {http://arxiv.org/abs/2303.02263} {arXiv:2303.02263 [astro-ph.GA]} \BibitemShut {NoStop}%
\bibitem [{\citenamefont {et~al}(2022{\natexlab{a}})}]{Vanzella}%
  \BibitemOpen
  \bibfield  {author} {\bibinfo {author} {\bibfnamefont {V.~E.}\ \bibnamefont {et~al}},\ }\href {\doibase 10.3847/2041-8213/ac8c2d} {\bibfield  {journal} {\bibinfo  {journal} {\apjl}\ }\textbf {\bibinfo {volume} {940}},\ \bibinfo {pages} {L53} (\bibinfo {year} {2022}{\natexlab{a}})},\ \Eprint {http://arxiv.org/abs/2208.00520} {arXiv:2208.00520 [astro-ph.GA]} \BibitemShut {NoStop}%
\bibitem [{\citenamefont {et~al}(2022{\natexlab{b}})}]{Lamiya}%
  \BibitemOpen
  \bibfield  {author} {\bibinfo {author} {\bibfnamefont {L.~M.}\ \bibnamefont {et~al}},\ }\href {\doibase 0.3847/2041-8213/ac90ca} {\bibfield  {journal} {\bibinfo  {journal} {\apjl}\ }\textbf {\bibinfo {volume} {937}},\ \bibinfo {pages} {L35} (\bibinfo {year} {2022}{\natexlab{b}})},\ \Eprint {http://arxiv.org/abs/2208.02233} {arXiv:2208.02233 [astro-ph.GA]} \BibitemShut {NoStop}%
\bibitem [{\citenamefont {Vitale}\ \emph {et~al.}(2019)\citenamefont {Vitale}, \citenamefont {Farr}, \citenamefont {Ng},\ and\ \citenamefont {Rodriguez}}]{Vitale:2018yhm}%
  \BibitemOpen
  \bibfield  {author} {\bibinfo {author} {\bibfnamefont {S.}~\bibnamefont {Vitale}}, \bibinfo {author} {\bibfnamefont {W.~M.}\ \bibnamefont {Farr}}, \bibinfo {author} {\bibfnamefont {K.}~\bibnamefont {Ng}}, \ and\ \bibinfo {author} {\bibfnamefont {C.~L.}\ \bibnamefont {Rodriguez}},\ }\href {\doibase 10.3847/2041-8213/ab50c0} {\bibfield  {journal} {\bibinfo  {journal} {Astrophys. J. Lett.}\ }\textbf {\bibinfo {volume} {886}},\ \bibinfo {pages} {L1} (\bibinfo {year} {2019})},\ \Eprint {http://arxiv.org/abs/1808.00901} {arXiv:1808.00901 [astro-ph.HE]} \BibitemShut {NoStop}%
\bibitem [{\citenamefont {Abbott}\ \emph {et~al.}(2018)\citenamefont {Abbott} \emph {et~al.}}]{KAGRA:2013rdx}%
  \BibitemOpen
  \bibfield  {author} {\bibinfo {author} {\bibfnamefont {B.~P.}\ \bibnamefont {Abbott}} \emph {et~al.} (\bibinfo {collaboration} {KAGRA, LIGO Scientific, Virgo, VIRGO}),\ }\href {\doibase 10.1007/s41114-020-00026-9} {\bibfield  {journal} {\bibinfo  {journal} {Living Rev. Rel.}\ }\textbf {\bibinfo {volume} {21}},\ \bibinfo {pages} {3} (\bibinfo {year} {2018})},\ \Eprint {http://arxiv.org/abs/1304.0670} {arXiv:1304.0670 [gr-qc]} \BibitemShut {NoStop}%
\bibitem [{\citenamefont {Romero-Shaw}\ \emph {et~al.}(2021)\citenamefont {Romero-Shaw}, \citenamefont {Kremer}, \citenamefont {Lasky}, \citenamefont {Thrane},\ and\ \citenamefont {Samsing}}]{Romero-Shaw:2020siz}%
  \BibitemOpen
  \bibfield  {author} {\bibinfo {author} {\bibfnamefont {I.~M.}\ \bibnamefont {Romero-Shaw}}, \bibinfo {author} {\bibfnamefont {K.}~\bibnamefont {Kremer}}, \bibinfo {author} {\bibfnamefont {P.~D.}\ \bibnamefont {Lasky}}, \bibinfo {author} {\bibfnamefont {E.}~\bibnamefont {Thrane}}, \ and\ \bibinfo {author} {\bibfnamefont {J.}~\bibnamefont {Samsing}},\ }\href {\doibase 10.1093/mnras/stab1815} {\bibfield  {journal} {\bibinfo  {journal} {Mon. Not. Roy. Astron. Soc.}\ }\textbf {\bibinfo {volume} {506}},\ \bibinfo {pages} {2362} (\bibinfo {year} {2021})},\ \Eprint {http://arxiv.org/abs/2011.14541} {arXiv:2011.14541 [astro-ph.HE]} \BibitemShut {NoStop}%
\bibitem [{\citenamefont {Abbott}\ \emph {et~al.}(2019{\natexlab{a}})\citenamefont {Abbott} \emph {et~al.}}]{LIGOScientific:2019vic}%
  \BibitemOpen
  \bibfield  {author} {\bibinfo {author} {\bibfnamefont {B.~P.}\ \bibnamefont {Abbott}} \emph {et~al.} (\bibinfo {collaboration} {LIGO Scientific, Virgo}),\ }\href {\doibase 10.1103/PhysRevD.100.061101} {\bibfield  {journal} {\bibinfo  {journal} {Phys. Rev. D}\ }\textbf {\bibinfo {volume} {100}},\ \bibinfo {pages} {061101} (\bibinfo {year} {2019}{\natexlab{a}})},\ \Eprint {http://arxiv.org/abs/1903.02886} {arXiv:1903.02886 [gr-qc]} \BibitemShut {NoStop}%
\bibitem [{\citenamefont {Abbott}\ \emph {et~al.}(2019{\natexlab{b}})\citenamefont {Abbott} \emph {et~al.}}]{LIGOScientific:2018mvr}%
  \BibitemOpen
  \bibfield  {author} {\bibinfo {author} {\bibfnamefont {B.~P.}\ \bibnamefont {Abbott}} \emph {et~al.} (\bibinfo {collaboration} {LIGO Scientific, Virgo}),\ }\href {\doibase 10.1103/PhysRevX.9.031040} {\bibfield  {journal} {\bibinfo  {journal} {Phys. Rev. X}\ }\textbf {\bibinfo {volume} {9}},\ \bibinfo {pages} {031040} (\bibinfo {year} {2019}{\natexlab{b}})},\ \Eprint {http://arxiv.org/abs/1811.12907} {arXiv:1811.12907 [astro-ph.HE]} \BibitemShut {NoStop}%
\bibitem [{\citenamefont {Abbott}\ \emph {et~al.}(2017)\citenamefont {Abbott} \emph {et~al.}}]{LIGOScientific:2016jlg}%
  \BibitemOpen
  \bibfield  {author} {\bibinfo {author} {\bibfnamefont {B.~P.}\ \bibnamefont {Abbott}} \emph {et~al.} (\bibinfo {collaboration} {LIGO Scientific, Virgo}),\ }\href {\doibase 10.1103/PhysRevLett.118.121101} {\bibfield  {journal} {\bibinfo  {journal} {Phys. Rev. Lett.}\ }\textbf {\bibinfo {volume} {118}},\ \bibinfo {pages} {121101} (\bibinfo {year} {2017})},\ \bibinfo {note} {[Erratum: Phys.Rev.Lett. 119, 029901 (2017)]},\ \Eprint {http://arxiv.org/abs/1612.02029} {arXiv:1612.02029 [gr-qc]} \BibitemShut {NoStop}%
\bibitem [{\citenamefont {Abbott}\ \emph {et~al.}(2023)\citenamefont {Abbott} \emph {et~al.}}]{KAGRA:2021duu}%
  \BibitemOpen
  \bibfield  {author} {\bibinfo {author} {\bibfnamefont {R.}~\bibnamefont {Abbott}} \emph {et~al.} (\bibinfo {collaboration} {KAGRA, VIRGO, LIGO Scientific}),\ }\href {\doibase 10.1103/PhysRevX.13.011048} {\bibfield  {journal} {\bibinfo  {journal} {Phys. Rev. X}\ }\textbf {\bibinfo {volume} {13}},\ \bibinfo {pages} {011048} (\bibinfo {year} {2023})},\ \Eprint {http://arxiv.org/abs/2111.03634} {arXiv:2111.03634 [astro-ph.HE]} \BibitemShut {NoStop}%
\bibitem [{\citenamefont {Zhao}\ and\ \citenamefont {Lu}(2020)}]{Zhao:2020iew}%
  \BibitemOpen
  \bibfield  {author} {\bibinfo {author} {\bibfnamefont {Y.}~\bibnamefont {Zhao}}\ and\ \bibinfo {author} {\bibfnamefont {Y.}~\bibnamefont {Lu}},\ }\href {\doibase 10.1093/mnras/staa2707} {\bibfield  {journal} {\bibinfo  {journal} {Mon. Not. Roy. Astron. Soc.}\ }\textbf {\bibinfo {volume} {500}},\ \bibinfo {pages} {1421} (\bibinfo {year} {2020})},\ \Eprint {http://arxiv.org/abs/2009.01436} {arXiv:2009.01436 [astro-ph.HE]} \BibitemShut {NoStop}%
\bibitem [{\citenamefont {P\'erigois}\ \emph {et~al.}(2022)\citenamefont {P\'erigois}, \citenamefont {Santoliquido}, \citenamefont {Bouffanais}, \citenamefont {Di~Carlo}, \citenamefont {Giacobbo}, \citenamefont {Rastello}, \citenamefont {Mapelli},\ and\ \citenamefont {Regimbau}}]{Perigois:2021ovr}%
  \BibitemOpen
  \bibfield  {author} {\bibinfo {author} {\bibfnamefont {C.}~\bibnamefont {P\'erigois}}, \bibinfo {author} {\bibfnamefont {F.}~\bibnamefont {Santoliquido}}, \bibinfo {author} {\bibfnamefont {Y.}~\bibnamefont {Bouffanais}}, \bibinfo {author} {\bibfnamefont {U.~N.}\ \bibnamefont {Di~Carlo}}, \bibinfo {author} {\bibfnamefont {N.}~\bibnamefont {Giacobbo}}, \bibinfo {author} {\bibfnamefont {S.}~\bibnamefont {Rastello}}, \bibinfo {author} {\bibfnamefont {M.}~\bibnamefont {Mapelli}}, \ and\ \bibinfo {author} {\bibfnamefont {T.}~\bibnamefont {Regimbau}},\ }\href {\doibase 10.1103/PhysRevD.105.103032} {\bibfield  {journal} {\bibinfo  {journal} {Phys. Rev. D}\ }\textbf {\bibinfo {volume} {105}},\ \bibinfo {pages} {103032} (\bibinfo {year} {2022})},\ \Eprint {http://arxiv.org/abs/2112.01119} {arXiv:2112.01119 [astro-ph.CO]} \BibitemShut {NoStop}%
\bibitem [{\citenamefont {Bavera}\ \emph {et~al.}(2022)\citenamefont {Bavera}, \citenamefont {Franciolini}, \citenamefont {Cusin}, \citenamefont {Riotto}, \citenamefont {Zevin},\ and\ \citenamefont {Fragos}}]{Bavera:2021wmw}%
  \BibitemOpen
  \bibfield  {author} {\bibinfo {author} {\bibfnamefont {S.~S.}\ \bibnamefont {Bavera}}, \bibinfo {author} {\bibfnamefont {G.}~\bibnamefont {Franciolini}}, \bibinfo {author} {\bibfnamefont {G.}~\bibnamefont {Cusin}}, \bibinfo {author} {\bibfnamefont {A.}~\bibnamefont {Riotto}}, \bibinfo {author} {\bibfnamefont {M.}~\bibnamefont {Zevin}}, \ and\ \bibinfo {author} {\bibfnamefont {T.}~\bibnamefont {Fragos}},\ }\href {\doibase 10.1051/0004-6361/202142208} {\bibfield  {journal} {\bibinfo  {journal} {Astron. Astrophys.}\ }\textbf {\bibinfo {volume} {660}},\ \bibinfo {pages} {A26} (\bibinfo {year} {2022})},\ \Eprint {http://arxiv.org/abs/2109.05836} {arXiv:2109.05836 [astro-ph.CO]} \BibitemShut {NoStop}%
\bibitem [{\citenamefont {Kremer}\ \emph {et~al.}(2020)\citenamefont {Kremer}, \citenamefont {Ye}, \citenamefont {Rui}, \citenamefont {Weatherford}, \citenamefont {Chatterjee}, \citenamefont {Fragione}, \citenamefont {Rodriguez}, \citenamefont {Spera},\ and\ \citenamefont {Rasio}}]{Kremer:2019iul}%
  \BibitemOpen
  \bibfield  {author} {\bibinfo {author} {\bibfnamefont {K.}~\bibnamefont {Kremer}}, \bibinfo {author} {\bibfnamefont {C.~S.}\ \bibnamefont {Ye}}, \bibinfo {author} {\bibfnamefont {N.~Z.}\ \bibnamefont {Rui}}, \bibinfo {author} {\bibfnamefont {N.~C.}\ \bibnamefont {Weatherford}}, \bibinfo {author} {\bibfnamefont {S.}~\bibnamefont {Chatterjee}}, \bibinfo {author} {\bibfnamefont {G.}~\bibnamefont {Fragione}}, \bibinfo {author} {\bibfnamefont {C.~L.}\ \bibnamefont {Rodriguez}}, \bibinfo {author} {\bibfnamefont {M.}~\bibnamefont {Spera}}, \ and\ \bibinfo {author} {\bibfnamefont {F.~A.}\ \bibnamefont {Rasio}},\ }\href {\doibase 10.3847/1538-4365/ab7919} {\bibfield  {journal} {\bibinfo  {journal} {Astrophys. J. Suppl.}\ }\textbf {\bibinfo {volume} {247}},\ \bibinfo {pages} {48} (\bibinfo {year} {2020})},\ \Eprint {http://arxiv.org/abs/1911.00018} {arXiv:1911.00018 [astro-ph.HE]} \BibitemShut {NoStop}%
\bibitem [{\citenamefont {Rodriguez}\ \emph {et~al.}(2022)\citenamefont {Rodriguez} \emph {et~al.}}]{Rodriguez:2021qhl}%
  \BibitemOpen
  \bibfield  {author} {\bibinfo {author} {\bibfnamefont {C.~L.}\ \bibnamefont {Rodriguez}} \emph {et~al.},\ }\href {\doibase 10.3847/1538-4365/ac2edf} {\bibfield  {journal} {\bibinfo  {journal} {Astrophys. J. Supp.}\ }\textbf {\bibinfo {volume} {258}},\ \bibinfo {pages} {22} (\bibinfo {year} {2022})},\ \Eprint {http://arxiv.org/abs/2106.02643} {arXiv:2106.02643 [astro-ph.GA]} \BibitemShut {NoStop}%
\bibitem [{\citenamefont {{King}}(1966)}]{King1966}%
  \BibitemOpen
  \bibfield  {author} {\bibinfo {author} {\bibfnamefont {I.~R.}\ \bibnamefont {{King}}},\ }\href {\doibase 10.1086/109857} {\bibfield  {journal} {\bibinfo  {journal} {\aj}\ }\textbf {\bibinfo {volume} {71}},\ \bibinfo {pages} {64} (\bibinfo {year} {1966})}\BibitemShut {NoStop}%
\bibitem [{\citenamefont {{Kroupa}}(2001)}]{Kroupa2001}%
  \BibitemOpen
  \bibfield  {author} {\bibinfo {author} {\bibfnamefont {P.}~\bibnamefont {{Kroupa}}},\ }\href {\doibase 10.1046/j.1365-8711.2001.04022.x} {\bibfield  {journal} {\bibinfo  {journal} {\mnras}\ }\textbf {\bibinfo {volume} {322}},\ \bibinfo {pages} {231} (\bibinfo {year} {2001})},\ \Eprint {http://arxiv.org/abs/astro-ph/0009005} {arXiv:astro-ph/0009005 [astro-ph]} \BibitemShut {NoStop}%
\bibitem [{\citenamefont {{Duquennoy}}\ and\ \citenamefont {{Mayor}}(1991)}]{DuquennoyMayor1991}%
  \BibitemOpen
  \bibfield  {author} {\bibinfo {author} {\bibfnamefont {A.}~\bibnamefont {{Duquennoy}}}\ and\ \bibinfo {author} {\bibfnamefont {M.}~\bibnamefont {{Mayor}}},\ }\href@noop {} {\bibfield  {journal} {\bibinfo  {journal} {\aap}\ }\textbf {\bibinfo {volume} {248}},\ \bibinfo {pages} {485} (\bibinfo {year} {1991})}\BibitemShut {NoStop}%
\bibitem [{\citenamefont {{Heggie}}(1975)}]{Heggie1975}%
  \BibitemOpen
  \bibfield  {author} {\bibinfo {author} {\bibfnamefont {D.~C.}\ \bibnamefont {{Heggie}}},\ }\href {\doibase 10.1093/mnras/173.3.729} {\bibfield  {journal} {\bibinfo  {journal} {\mnras}\ }\textbf {\bibinfo {volume} {173}},\ \bibinfo {pages} {729} (\bibinfo {year} {1975})}\BibitemShut {NoStop}%
\bibitem [{\citenamefont {{Hurley}}\ \emph {et~al.}(2000)\citenamefont {{Hurley}}, \citenamefont {{Pols}},\ and\ \citenamefont {{Tout}}}]{HurleyPols2000}%
  \BibitemOpen
  \bibfield  {author} {\bibinfo {author} {\bibfnamefont {J.~R.}\ \bibnamefont {{Hurley}}}, \bibinfo {author} {\bibfnamefont {O.~R.}\ \bibnamefont {{Pols}}}, \ and\ \bibinfo {author} {\bibfnamefont {C.~A.}\ \bibnamefont {{Tout}}},\ }\href {\doibase 10.1046/j.1365-8711.2000.03426.x} {\bibfield  {journal} {\bibinfo  {journal} {\mnras}\ }\textbf {\bibinfo {volume} {315}},\ \bibinfo {pages} {543} (\bibinfo {year} {2000})},\ \Eprint {http://arxiv.org/abs/astro-ph/0001295} {arXiv:astro-ph/0001295 [astro-ph]} \BibitemShut {NoStop}%
\bibitem [{\citenamefont {{Hurley}}\ \emph {et~al.}(2002)\citenamefont {{Hurley}}, \citenamefont {{Tout}},\ and\ \citenamefont {{Pols}}}]{HurleyTout2002}%
  \BibitemOpen
  \bibfield  {author} {\bibinfo {author} {\bibfnamefont {J.~R.}\ \bibnamefont {{Hurley}}}, \bibinfo {author} {\bibfnamefont {C.~A.}\ \bibnamefont {{Tout}}}, \ and\ \bibinfo {author} {\bibfnamefont {O.~R.}\ \bibnamefont {{Pols}}},\ }\href {\doibase 10.1046/j.1365-8711.2002.05038.x} {\bibfield  {journal} {\bibinfo  {journal} {\mnras}\ }\textbf {\bibinfo {volume} {329}},\ \bibinfo {pages} {897} (\bibinfo {year} {2002})},\ \Eprint {http://arxiv.org/abs/astro-ph/0201220} {arXiv:astro-ph/0201220 [astro-ph]} \BibitemShut {NoStop}%
\bibitem [{\citenamefont {{Fryer}}\ \emph {et~al.}(2012)\citenamefont {{Fryer}}, \citenamefont {{Belczynski}}, \citenamefont {{Wiktorowicz}}, \citenamefont {{Dominik}},\ and\ \citenamefont {et~al.}}]{FryerBelczynski2012}%
  \BibitemOpen
  \bibfield  {author} {\bibinfo {author} {\bibfnamefont {C.~L.}\ \bibnamefont {{Fryer}}}, \bibinfo {author} {\bibfnamefont {K.}~\bibnamefont {{Belczynski}}}, \bibinfo {author} {\bibfnamefont {G.}~\bibnamefont {{Wiktorowicz}}}, \bibinfo {author} {\bibfnamefont {M.}~\bibnamefont {{Dominik}}}, \ and\ \bibinfo {author} {\bibnamefont {et~al.}},\ }\href {\doibase 10.1088/0004-637X/749/1/91} {\bibfield  {journal} {\bibinfo  {journal} {\apj}\ }\textbf {\bibinfo {volume} {749}},\ \bibinfo {eid} {91} (\bibinfo {year} {2012})},\ \Eprint {http://arxiv.org/abs/1110.1726} {arXiv:1110.1726 [astro-ph.SR]} \BibitemShut {NoStop}%
\bibitem [{\citenamefont {{Hobbs}}\ \emph {et~al.}(2005)\citenamefont {{Hobbs}}, \citenamefont {{Lorimer}}, \citenamefont {{Lyne}},\ and\ \citenamefont {{Kramer}}}]{HobbsLorimer2005}%
  \BibitemOpen
  \bibfield  {author} {\bibinfo {author} {\bibfnamefont {G.}~\bibnamefont {{Hobbs}}}, \bibinfo {author} {\bibfnamefont {D.~R.}\ \bibnamefont {{Lorimer}}}, \bibinfo {author} {\bibfnamefont {A.~G.}\ \bibnamefont {{Lyne}}}, \ and\ \bibinfo {author} {\bibfnamefont {M.}~\bibnamefont {{Kramer}}},\ }\href {\doibase 10.1111/j.1365-2966.2005.09087.x} {\bibfield  {journal} {\bibinfo  {journal} {\mnras}\ }\textbf {\bibinfo {volume} {360}},\ \bibinfo {pages} {974} (\bibinfo {year} {2005})},\ \Eprint {http://arxiv.org/abs/astro-ph/0504584} {arXiv:astro-ph/0504584 [astro-ph]} \BibitemShut {NoStop}%
\bibitem [{\citenamefont {Madau}\ and\ \citenamefont {Dickinson}(2014)}]{Madau:2014bja}%
  \BibitemOpen
  \bibfield  {author} {\bibinfo {author} {\bibfnamefont {P.}~\bibnamefont {Madau}}\ and\ \bibinfo {author} {\bibfnamefont {M.}~\bibnamefont {Dickinson}},\ }\href {\doibase 10.1146/annurev-astro-081811-125615} {\bibfield  {journal} {\bibinfo  {journal} {Ann. Rev. Astron. Astrophys.}\ }\textbf {\bibinfo {volume} {52}},\ \bibinfo {pages} {415} (\bibinfo {year} {2014})},\ \Eprint {http://arxiv.org/abs/1403.0007} {arXiv:1403.0007 [astro-ph.CO]} \BibitemShut {NoStop}%
\bibitem [{\citenamefont {Rodriguez}\ and\ \citenamefont {Loeb}(2018)}]{Rodriguez:2018rmd}%
  \BibitemOpen
  \bibfield  {author} {\bibinfo {author} {\bibfnamefont {C.~L.}\ \bibnamefont {Rodriguez}}\ and\ \bibinfo {author} {\bibfnamefont {A.}~\bibnamefont {Loeb}},\ }\href {\doibase 10.3847/2041-8213/aae377} {\bibfield  {journal} {\bibinfo  {journal} {Astrophys. J. Lett.}\ }\textbf {\bibinfo {volume} {866}},\ \bibinfo {pages} {L5} (\bibinfo {year} {2018})},\ \Eprint {http://arxiv.org/abs/1809.01152} {arXiv:1809.01152 [astro-ph.HE]} \BibitemShut {NoStop}%
\bibitem [{\citenamefont {Fragione}\ and\ \citenamefont {Kocsis}(2018)}]{Fragione:2018vty}%
  \BibitemOpen
  \bibfield  {author} {\bibinfo {author} {\bibfnamefont {G.}~\bibnamefont {Fragione}}\ and\ \bibinfo {author} {\bibfnamefont {B.}~\bibnamefont {Kocsis}},\ }\href {\doibase 10.1103/PhysRevLett.121.161103} {\bibfield  {journal} {\bibinfo  {journal} {Phys. Rev. Lett.}\ }\textbf {\bibinfo {volume} {121}},\ \bibinfo {pages} {161103} (\bibinfo {year} {2018})},\ \Eprint {http://arxiv.org/abs/1806.02351} {arXiv:1806.02351 [astro-ph.GA]} \BibitemShut {NoStop}%
\bibitem [{\citenamefont {Choksi}\ \emph {et~al.}(2019)\citenamefont {Choksi}, \citenamefont {Volonteri}, \citenamefont {Colpi}, \citenamefont {Gnedin},\ and\ \citenamefont {Li}}]{Choksi:2018jnq}%
  \BibitemOpen
  \bibfield  {author} {\bibinfo {author} {\bibfnamefont {N.}~\bibnamefont {Choksi}}, \bibinfo {author} {\bibfnamefont {M.}~\bibnamefont {Volonteri}}, \bibinfo {author} {\bibfnamefont {M.}~\bibnamefont {Colpi}}, \bibinfo {author} {\bibfnamefont {O.~Y.}\ \bibnamefont {Gnedin}}, \ and\ \bibinfo {author} {\bibfnamefont {H.}~\bibnamefont {Li}},\ }\href {\doibase 10.3847/1538-4357/aaffde} {\bibfield  {journal} {\bibinfo  {journal} {Astrophys. J.}\ }\textbf {\bibinfo {volume} {873}},\ \bibinfo {pages} {100} (\bibinfo {year} {2019})},\ \Eprint {http://arxiv.org/abs/1809.01164} {arXiv:1809.01164 [astro-ph.GA]} \BibitemShut {NoStop}%
\bibitem [{\citenamefont {Antonini}\ and\ \citenamefont {Gieles}(2020{\natexlab{a}})}]{Antonini:2019ulv}%
  \BibitemOpen
  \bibfield  {author} {\bibinfo {author} {\bibfnamefont {F.}~\bibnamefont {Antonini}}\ and\ \bibinfo {author} {\bibfnamefont {M.}~\bibnamefont {Gieles}},\ }\href {\doibase 10.1093/mnras/stz3584} {\bibfield  {journal} {\bibinfo  {journal} {Mon. Not. Roy. Astron. Soc.}\ }\textbf {\bibinfo {volume} {492}},\ \bibinfo {pages} {2936} (\bibinfo {year} {2020}{\natexlab{a}})},\ \Eprint {http://arxiv.org/abs/1906.11855} {arXiv:1906.11855 [astro-ph.HE]} \BibitemShut {NoStop}%
\bibitem [{\citenamefont {Portegies~Zwart}\ and\ \citenamefont {McMillan}(2000)}]{PortegiesZwart:1999nm}%
  \BibitemOpen
  \bibfield  {author} {\bibinfo {author} {\bibfnamefont {S.~F.}\ \bibnamefont {Portegies~Zwart}}\ and\ \bibinfo {author} {\bibfnamefont {S.}~\bibnamefont {McMillan}},\ }\href {\doibase 10.1086/312422} {\bibfield  {journal} {\bibinfo  {journal} {Astrophys. J. Lett.}\ }\textbf {\bibinfo {volume} {528}},\ \bibinfo {pages} {L17} (\bibinfo {year} {2000})},\ \Eprint {http://arxiv.org/abs/astro-ph/9910061} {arXiv:astro-ph/9910061} \BibitemShut {NoStop}%
\bibitem [{\citenamefont {Antonini}\ \emph {et~al.}(2023)\citenamefont {Antonini}, \citenamefont {Gieles}, \citenamefont {Dosopoulou},\ and\ \citenamefont {Chattopadhyay}}]{Antonini:2022vib}%
  \BibitemOpen
  \bibfield  {author} {\bibinfo {author} {\bibfnamefont {F.}~\bibnamefont {Antonini}}, \bibinfo {author} {\bibfnamefont {M.}~\bibnamefont {Gieles}}, \bibinfo {author} {\bibfnamefont {F.}~\bibnamefont {Dosopoulou}}, \ and\ \bibinfo {author} {\bibfnamefont {D.}~\bibnamefont {Chattopadhyay}},\ }\href {\doibase 10.1093/mnras/stad972} {\bibfield  {journal} {\bibinfo  {journal} {Mon. Not. Roy. Astron. Soc.}\ }\textbf {\bibinfo {volume} {522}},\ \bibinfo {pages} {466} (\bibinfo {year} {2023})},\ \Eprint {http://arxiv.org/abs/2208.01081} {arXiv:2208.01081 [astro-ph.HE]} \BibitemShut {NoStop}%
\bibitem [{\citenamefont {Madau}\ and\ \citenamefont {Fragos}(2017)}]{Madau:2016jbv}%
  \BibitemOpen
  \bibfield  {author} {\bibinfo {author} {\bibfnamefont {P.}~\bibnamefont {Madau}}\ and\ \bibinfo {author} {\bibfnamefont {T.}~\bibnamefont {Fragos}},\ }\href {\doibase 10.3847/1538-4357/aa6af9} {\bibfield  {journal} {\bibinfo  {journal} {Astrophys. J.}\ }\textbf {\bibinfo {volume} {840}},\ \bibinfo {pages} {39} (\bibinfo {year} {2017})},\ \Eprint {http://arxiv.org/abs/1606.07887} {arXiv:1606.07887 [astro-ph.GA]} \BibitemShut {NoStop}%
\bibitem [{\citenamefont {Harris}\ \emph {et~al.}(2013)\citenamefont {Harris}, \citenamefont {Harris},\ and\ \citenamefont {Alessi}}]{harris2013catalog}%
  \BibitemOpen
  \bibfield  {author} {\bibinfo {author} {\bibfnamefont {W.~E.}\ \bibnamefont {Harris}}, \bibinfo {author} {\bibfnamefont {G.~L.}\ \bibnamefont {Harris}}, \ and\ \bibinfo {author} {\bibfnamefont {M.}~\bibnamefont {Alessi}},\ }\href@noop {} {\bibfield  {journal} {\bibinfo  {journal} {The Astrophysical Journal}\ }\textbf {\bibinfo {volume} {772}},\ \bibinfo {pages} {82} (\bibinfo {year} {2013})}\BibitemShut {NoStop}%
\bibitem [{\citenamefont {Antonini}\ and\ \citenamefont {Gieles}(2020{\natexlab{b}})}]{Antonini:2020xnd}%
  \BibitemOpen
  \bibfield  {author} {\bibinfo {author} {\bibfnamefont {F.}~\bibnamefont {Antonini}}\ and\ \bibinfo {author} {\bibfnamefont {M.}~\bibnamefont {Gieles}},\ }\href {\doibase 10.1103/PhysRevD.102.123016} {\bibfield  {journal} {\bibinfo  {journal} {Phys. Rev. D}\ }\textbf {\bibinfo {volume} {102}},\ \bibinfo {pages} {123016} (\bibinfo {year} {2020}{\natexlab{b}})},\ \Eprint {http://arxiv.org/abs/2009.01861} {arXiv:2009.01861 [astro-ph.HE]} \BibitemShut {NoStop}%
\bibitem [{\citenamefont {Gnedin}\ \emph {et~al.}(2014)\citenamefont {Gnedin}, \citenamefont {Ostriker},\ and\ \citenamefont {Tremaine}}]{Gnedin:2013cda}%
  \BibitemOpen
  \bibfield  {author} {\bibinfo {author} {\bibfnamefont {O.~Y.}\ \bibnamefont {Gnedin}}, \bibinfo {author} {\bibfnamefont {J.~P.}\ \bibnamefont {Ostriker}}, \ and\ \bibinfo {author} {\bibfnamefont {S.}~\bibnamefont {Tremaine}},\ }\href {\doibase 10.1088/0004-637X/785/1/71} {\bibfield  {journal} {\bibinfo  {journal} {Astrophys. J.}\ }\textbf {\bibinfo {volume} {785}},\ \bibinfo {pages} {71} (\bibinfo {year} {2014})},\ \Eprint {http://arxiv.org/abs/1308.0021} {arXiv:1308.0021 [astro-ph.CO]} \BibitemShut {NoStop}%
\bibitem [{\citenamefont {Allen}\ and\ \citenamefont {Romano}(1999)}]{Allen:1997ad}%
  \BibitemOpen
  \bibfield  {author} {\bibinfo {author} {\bibfnamefont {B.}~\bibnamefont {Allen}}\ and\ \bibinfo {author} {\bibfnamefont {J.~D.}\ \bibnamefont {Romano}},\ }\href {\doibase 10.1103/PhysRevD.59.102001} {\bibfield  {journal} {\bibinfo  {journal} {Phys. Rev. D}\ }\textbf {\bibinfo {volume} {59}},\ \bibinfo {pages} {102001} (\bibinfo {year} {1999})},\ \Eprint {http://arxiv.org/abs/gr-qc/9710117} {arXiv:gr-qc/9710117} \BibitemShut {NoStop}%
\bibitem [{\citenamefont {Allen}(1996)}]{Allen:1996vm}%
  \BibitemOpen
  \bibfield  {author} {\bibinfo {author} {\bibfnamefont {B.}~\bibnamefont {Allen}},\ }in\ \href@noop {} {\emph {\bibinfo {booktitle} {{Les Houches School of Physics: Astrophysical Sources of Gravitational Radiation}}}}\ (\bibinfo {year} {1996})\ pp.\ \bibinfo {pages} {373--417},\ \Eprint {http://arxiv.org/abs/gr-qc/9604033} {arXiv:gr-qc/9604033} \BibitemShut {NoStop}%
\bibitem [{\citenamefont {Phinney}(2001)}]{Phinney:2001di}%
  \BibitemOpen
  \bibfield  {author} {\bibinfo {author} {\bibfnamefont {E.~S.}\ \bibnamefont {Phinney}},\ }\href@noop {} {\  (\bibinfo {year} {2001})},\ \Eprint {http://arxiv.org/abs/astro-ph/0108028} {arXiv:astro-ph/0108028} \BibitemShut {NoStop}%
\bibitem [{\citenamefont {Romano}\ and\ \citenamefont {Cornish}(2017)}]{Romano:2016dpx}%
  \BibitemOpen
  \bibfield  {author} {\bibinfo {author} {\bibfnamefont {J.~D.}\ \bibnamefont {Romano}}\ and\ \bibinfo {author} {\bibfnamefont {N.~J.}\ \bibnamefont {Cornish}},\ }\href {\doibase 10.1007/s41114-017-0004-1} {\bibfield  {journal} {\bibinfo  {journal} {Living Rev. Rel.}\ }\textbf {\bibinfo {volume} {20}},\ \bibinfo {pages} {2} (\bibinfo {year} {2017})},\ \Eprint {http://arxiv.org/abs/1608.06889} {arXiv:1608.06889 [gr-qc]} \BibitemShut {NoStop}%
\bibitem [{\citenamefont {Aghanim}\ \emph {et~al.}(2020)\citenamefont {Aghanim} \emph {et~al.}}]{Planck:2018vyg}%
  \BibitemOpen
  \bibfield  {author} {\bibinfo {author} {\bibfnamefont {N.}~\bibnamefont {Aghanim}} \emph {et~al.} (\bibinfo {collaboration} {Planck}),\ }\href {\doibase 10.1051/0004-6361/201833910} {\bibfield  {journal} {\bibinfo  {journal} {Astron. Astrophys.}\ }\textbf {\bibinfo {volume} {641}},\ \bibinfo {pages} {A6} (\bibinfo {year} {2020})},\ \bibinfo {note} {[Erratum: Astron.Astrophys. 652, C4 (2021)]},\ \Eprint {http://arxiv.org/abs/1807.06209} {arXiv:1807.06209 [astro-ph.CO]} \BibitemShut {NoStop}%
\bibitem [{\citenamefont {Ajith}\ \emph {et~al.}(2011)\citenamefont {Ajith} \emph {et~al.}}]{Ajith:2009bn}%
  \BibitemOpen
  \bibfield  {author} {\bibinfo {author} {\bibfnamefont {P.}~\bibnamefont {Ajith}} \emph {et~al.},\ }\href {\doibase 10.1103/PhysRevLett.106.241101} {\bibfield  {journal} {\bibinfo  {journal} {Phys. Rev. Lett.}\ }\textbf {\bibinfo {volume} {106}},\ \bibinfo {pages} {241101} (\bibinfo {year} {2011})},\ \Eprint {http://arxiv.org/abs/0909.2867} {arXiv:0909.2867 [gr-qc]} \BibitemShut {NoStop}%
\bibitem [{\citenamefont {Zhu}\ \emph {et~al.}(2011)\citenamefont {Zhu}, \citenamefont {Howell}, \citenamefont {Regimbau}, \citenamefont {Blair},\ and\ \citenamefont {Zhu}}]{Zhu:2011bd}%
  \BibitemOpen
  \bibfield  {author} {\bibinfo {author} {\bibfnamefont {X.-J.}\ \bibnamefont {Zhu}}, \bibinfo {author} {\bibfnamefont {E.}~\bibnamefont {Howell}}, \bibinfo {author} {\bibfnamefont {T.}~\bibnamefont {Regimbau}}, \bibinfo {author} {\bibfnamefont {D.}~\bibnamefont {Blair}}, \ and\ \bibinfo {author} {\bibfnamefont {Z.-H.}\ \bibnamefont {Zhu}},\ }\href {\doibase 10.1088/0004-637X/739/2/86} {\bibfield  {journal} {\bibinfo  {journal} {Astrophys. J.}\ }\textbf {\bibinfo {volume} {739}},\ \bibinfo {pages} {86} (\bibinfo {year} {2011})},\ \Eprint {http://arxiv.org/abs/1104.3565} {arXiv:1104.3565 [gr-qc]} \BibitemShut {NoStop}%
\bibitem [{\citenamefont {Talbot}\ and\ \citenamefont {Thrane}(2018)}]{Talbot:2018cva}%
  \BibitemOpen
  \bibfield  {author} {\bibinfo {author} {\bibfnamefont {C.}~\bibnamefont {Talbot}}\ and\ \bibinfo {author} {\bibfnamefont {E.}~\bibnamefont {Thrane}},\ }\href {\doibase 10.3847/1538-4357/aab34c} {\bibfield  {journal} {\bibinfo  {journal} {Astrophys. J.}\ }\textbf {\bibinfo {volume} {856}},\ \bibinfo {pages} {173} (\bibinfo {year} {2018})},\ \Eprint {http://arxiv.org/abs/1801.02699} {arXiv:1801.02699 [astro-ph.HE]} \BibitemShut {NoStop}%
\bibitem [{\citenamefont {Callister}\ \emph {et~al.}(2020)\citenamefont {Callister}, \citenamefont {Fishbach}, \citenamefont {Holz},\ and\ \citenamefont {Farr}}]{Callister:2020arv}%
  \BibitemOpen
  \bibfield  {author} {\bibinfo {author} {\bibfnamefont {T.}~\bibnamefont {Callister}}, \bibinfo {author} {\bibfnamefont {M.}~\bibnamefont {Fishbach}}, \bibinfo {author} {\bibfnamefont {D.}~\bibnamefont {Holz}}, \ and\ \bibinfo {author} {\bibfnamefont {W.}~\bibnamefont {Farr}},\ }\href {\doibase 10.3847/2041-8213/ab9743} {\bibfield  {journal} {\bibinfo  {journal} {Astrophys. J. Lett.}\ }\textbf {\bibinfo {volume} {896}},\ \bibinfo {pages} {L32} (\bibinfo {year} {2020})},\ \Eprint {http://arxiv.org/abs/2003.12152} {arXiv:2003.12152 [astro-ph.HE]} \BibitemShut {NoStop}%
\bibitem [{\citenamefont {Thrane}\ and\ \citenamefont {Romano}(2013)}]{Thrane:2013oya}%
  \BibitemOpen
  \bibfield  {author} {\bibinfo {author} {\bibfnamefont {E.}~\bibnamefont {Thrane}}\ and\ \bibinfo {author} {\bibfnamefont {J.~D.}\ \bibnamefont {Romano}},\ }\href {\doibase 10.1103/PhysRevD.88.124032} {\bibfield  {journal} {\bibinfo  {journal} {Phys. Rev. D}\ }\textbf {\bibinfo {volume} {88}},\ \bibinfo {pages} {124032} (\bibinfo {year} {2013})},\ \Eprint {http://arxiv.org/abs/1310.5300} {arXiv:1310.5300 [astro-ph.IM]} \BibitemShut {NoStop}%
\bibitem [{\citenamefont {Lehoucq}\ \emph {et~al.}(2023)\citenamefont {Lehoucq}, \citenamefont {Dvorkin}, \citenamefont {Srinivasan}, \citenamefont {Pellouin},\ and\ \citenamefont {Lamberts}}]{Lehoucq:2023zlt}%
  \BibitemOpen
  \bibfield  {author} {\bibinfo {author} {\bibfnamefont {L.}~\bibnamefont {Lehoucq}}, \bibinfo {author} {\bibfnamefont {I.}~\bibnamefont {Dvorkin}}, \bibinfo {author} {\bibfnamefont {R.}~\bibnamefont {Srinivasan}}, \bibinfo {author} {\bibfnamefont {C.}~\bibnamefont {Pellouin}}, \ and\ \bibinfo {author} {\bibfnamefont {A.}~\bibnamefont {Lamberts}},\ }\href@noop {} {\  (\bibinfo {year} {2023})},\ \Eprint {http://arxiv.org/abs/2306.09861} {arXiv:2306.09861 [astro-ph.HE]} \BibitemShut {NoStop}%
\bibitem [{\citenamefont {Turbang}\ \emph {et~al.}(2023)\citenamefont {Turbang}, \citenamefont {Lalleman}, \citenamefont {Callister},\ and\ \citenamefont {van Remortel}}]{Turbang:2023tjk}%
  \BibitemOpen
  \bibfield  {author} {\bibinfo {author} {\bibfnamefont {K.}~\bibnamefont {Turbang}}, \bibinfo {author} {\bibfnamefont {M.}~\bibnamefont {Lalleman}}, \bibinfo {author} {\bibfnamefont {T.~A.}\ \bibnamefont {Callister}}, \ and\ \bibinfo {author} {\bibfnamefont {N.}~\bibnamefont {van Remortel}},\ }\href@noop {} {\  (\bibinfo {year} {2023})},\ \Eprint {http://arxiv.org/abs/2310.17625} {arXiv:2310.17625 [astro-ph.HE]} \BibitemShut {NoStop}%
\bibitem [{\citenamefont {Christensen}(1992)}]{Christensen:1992wi}%
  \BibitemOpen
  \bibfield  {author} {\bibinfo {author} {\bibfnamefont {N.}~\bibnamefont {Christensen}},\ }\href {\doibase 10.1103/PhysRevD.46.5250} {\bibfield  {journal} {\bibinfo  {journal} {Phys. Rev. D}\ }\textbf {\bibinfo {volume} {46}},\ \bibinfo {pages} {5250} (\bibinfo {year} {1992})}\BibitemShut {NoStop}%
\bibitem [{\citenamefont {Flanagan}(1993)}]{Flanagan:1993ix}%
  \BibitemOpen
  \bibfield  {author} {\bibinfo {author} {\bibfnamefont {E.~E.}\ \bibnamefont {Flanagan}},\ }\href {\doibase 10.1103/PhysRevD.48.2389} {\bibfield  {journal} {\bibinfo  {journal} {Phys. Rev. D}\ }\textbf {\bibinfo {volume} {48}},\ \bibinfo {pages} {2389} (\bibinfo {year} {1993})},\ \Eprint {http://arxiv.org/abs/astro-ph/9305029} {arXiv:astro-ph/9305029} \BibitemShut {NoStop}%
\bibitem [{\citenamefont {Callister}\ \emph {et~al.}(2016)\citenamefont {Callister}, \citenamefont {Sammut}, \citenamefont {Qiu}, \citenamefont {Mandel},\ and\ \citenamefont {Thrane}}]{Callister:2016ewt}%
  \BibitemOpen
  \bibfield  {author} {\bibinfo {author} {\bibfnamefont {T.}~\bibnamefont {Callister}}, \bibinfo {author} {\bibfnamefont {L.}~\bibnamefont {Sammut}}, \bibinfo {author} {\bibfnamefont {S.}~\bibnamefont {Qiu}}, \bibinfo {author} {\bibfnamefont {I.}~\bibnamefont {Mandel}}, \ and\ \bibinfo {author} {\bibfnamefont {E.}~\bibnamefont {Thrane}},\ }\href {\doibase 10.1103/PhysRevX.6.031018} {\bibfield  {journal} {\bibinfo  {journal} {Phys. Rev. X}\ }\textbf {\bibinfo {volume} {6}},\ \bibinfo {pages} {031018} (\bibinfo {year} {2016})},\ \Eprint {http://arxiv.org/abs/1604.02513} {arXiv:1604.02513 [gr-qc]} \BibitemShut {NoStop}%
\bibitem [{\citenamefont {Fritschel}\ \emph {et~al.}()\citenamefont {Fritschel}, \citenamefont {Kuns}, \citenamefont {Driggers}, \citenamefont {Effler}, \citenamefont {Lantz}, \citenamefont {Ottaway}, \citenamefont {Ballmer}, \citenamefont {Dooley}, \citenamefont {Adhikari}, \citenamefont {Evans}, \citenamefont {Farr}, \citenamefont {Gonzalez}, \citenamefont {Schmidt},\ and\ \citenamefont {Raja}}]{T2200287}%
  \BibitemOpen
  \bibfield  {author} {\bibinfo {author} {\bibfnamefont {P.}~\bibnamefont {Fritschel}}, \bibinfo {author} {\bibfnamefont {K.}~\bibnamefont {Kuns}}, \bibinfo {author} {\bibfnamefont {J.}~\bibnamefont {Driggers}}, \bibinfo {author} {\bibfnamefont {A.}~\bibnamefont {Effler}}, \bibinfo {author} {\bibfnamefont {B.}~\bibnamefont {Lantz}}, \bibinfo {author} {\bibfnamefont {D.}~\bibnamefont {Ottaway}}, \bibinfo {author} {\bibfnamefont {S.}~\bibnamefont {Ballmer}}, \bibinfo {author} {\bibfnamefont {K.}~\bibnamefont {Dooley}}, \bibinfo {author} {\bibfnamefont {R.}~\bibnamefont {Adhikari}}, \bibinfo {author} {\bibfnamefont {M.}~\bibnamefont {Evans}}, \bibinfo {author} {\bibfnamefont {B.}~\bibnamefont {Farr}}, \bibinfo {author} {\bibfnamefont {G.}~\bibnamefont {Gonzalez}}, \bibinfo {author} {\bibfnamefont {P.}~\bibnamefont {Schmidt}}, \ and\ \bibinfo {author} {\bibfnamefont {S.}~\bibnamefont {Raja}},\ }\href@noop {} {\bibinfo  {journal} {Tech. Rep. T2200287 (LIGO, 2022)}\ }\BibitemShut {NoStop}%
\bibitem [{\citenamefont {Gupta}\ \emph {et~al.}(2023)\citenamefont {Gupta} \emph {et~al.}}]{Gupta:2023lga}%
  \BibitemOpen
\bibfield  {journal} {  }\bibfield  {author} {\bibinfo {author} {\bibfnamefont {I.}~\bibnamefont {Gupta}} \emph {et~al.},\ }\href@noop {} {\  (\bibinfo {year} {2023})},\ \Eprint {http://arxiv.org/abs/2307.10421} {arXiv:2307.10421 [gr-qc]} \BibitemShut {NoStop}%
\bibitem [{\citenamefont {Mukherjee}\ and\ \citenamefont {Silk}(2021)}]{Mukherjee:2021ags}%
  \BibitemOpen
  \bibfield  {author} {\bibinfo {author} {\bibfnamefont {S.}~\bibnamefont {Mukherjee}}\ and\ \bibinfo {author} {\bibfnamefont {J.}~\bibnamefont {Silk}},\ }\href {\doibase 10.1093/mnras/stab1932} {\bibfield  {journal} {\bibinfo  {journal} {Mon. Not. Roy. Astron. Soc.}\ }\textbf {\bibinfo {volume} {506}},\ \bibinfo {pages} {3977} (\bibinfo {year} {2021})},\ \Eprint {http://arxiv.org/abs/2105.11139} {arXiv:2105.11139 [gr-qc]} \BibitemShut {NoStop}%
\bibitem [{\citenamefont {Atal}\ \emph {et~al.}(2022)\citenamefont {Atal}, \citenamefont {Blanco-Pillado}, \citenamefont {Sanglas},\ and\ \citenamefont {Triantafyllou}}]{Atal:2022zux}%
  \BibitemOpen
  \bibfield  {author} {\bibinfo {author} {\bibfnamefont {V.}~\bibnamefont {Atal}}, \bibinfo {author} {\bibfnamefont {J.~J.}\ \bibnamefont {Blanco-Pillado}}, \bibinfo {author} {\bibfnamefont {A.}~\bibnamefont {Sanglas}}, \ and\ \bibinfo {author} {\bibfnamefont {N.}~\bibnamefont {Triantafyllou}},\ }\href {\doibase 10.1103/PhysRevD.105.123522} {\bibfield  {journal} {\bibinfo  {journal} {Phys. Rev. D}\ }\textbf {\bibinfo {volume} {105}},\ \bibinfo {pages} {123522} (\bibinfo {year} {2022})},\ \Eprint {http://arxiv.org/abs/2201.12218} {arXiv:2201.12218 [astro-ph.CO]} \BibitemShut {NoStop}%
\bibitem [{\citenamefont {{Vanzella}}\ \emph {et~al.}(2022)\citenamefont {{Vanzella}}, \citenamefont {{Castellano}}, \citenamefont {{Bergamini}}, \citenamefont {{Treu}},\ and\ \citenamefont {et~al.}}]{VanzellaCastellano2022}%
  \BibitemOpen
  \bibfield  {author} {\bibinfo {author} {\bibfnamefont {E.}~\bibnamefont {{Vanzella}}}, \bibinfo {author} {\bibfnamefont {M.}~\bibnamefont {{Castellano}}}, \bibinfo {author} {\bibfnamefont {P.}~\bibnamefont {{Bergamini}}}, \bibinfo {author} {\bibfnamefont {T.}~\bibnamefont {{Treu}}}, \ and\ \bibinfo {author} {\bibnamefont {et~al.}},\ }\href {\doibase 10.3847/2041-8213/ac8c2d} {\bibfield  {journal} {\bibinfo  {journal} {\apjl}\ }\textbf {\bibinfo {volume} {940}},\ \bibinfo {eid} {L53} (\bibinfo {year} {2022})},\ \Eprint {http://arxiv.org/abs/2208.00520} {arXiv:2208.00520 [astro-ph.GA]} \BibitemShut {NoStop}%
\bibitem [{\citenamefont {{Vanzella}}\ \emph {et~al.}(2023)\citenamefont {{Vanzella}}, \citenamefont {{Claeyssens}}, \citenamefont {{Welch}}, \citenamefont {{Adamo}},\ and\ \citenamefont {et~al.}}]{VanzellaClaeyssens2023}%
  \BibitemOpen
  \bibfield  {author} {\bibinfo {author} {\bibfnamefont {E.}~\bibnamefont {{Vanzella}}}, \bibinfo {author} {\bibfnamefont {A.}~\bibnamefont {{Claeyssens}}}, \bibinfo {author} {\bibfnamefont {B.}~\bibnamefont {{Welch}}}, \bibinfo {author} {\bibfnamefont {A.}~\bibnamefont {{Adamo}}}, \ and\ \bibinfo {author} {\bibnamefont {et~al.}},\ }\href {\doibase 10.3847/1538-4357/acb59a} {\bibfield  {journal} {\bibinfo  {journal} {\apj}\ }\textbf {\bibinfo {volume} {945}},\ \bibinfo {eid} {53} (\bibinfo {year} {2023})},\ \Eprint {http://arxiv.org/abs/2211.09839} {arXiv:2211.09839 [astro-ph.GA]} \BibitemShut {NoStop}%
\bibitem [{\citenamefont {Smith}\ and\ \citenamefont {Thrane}(2018)}]{Smith:2017vfk}%
  \BibitemOpen
  \bibfield  {author} {\bibinfo {author} {\bibfnamefont {R.}~\bibnamefont {Smith}}\ and\ \bibinfo {author} {\bibfnamefont {E.}~\bibnamefont {Thrane}},\ }\href {\doibase 10.1103/PhysRevX.8.021019} {\bibfield  {journal} {\bibinfo  {journal} {Phys. Rev. X}\ }\textbf {\bibinfo {volume} {8}},\ \bibinfo {pages} {021019} (\bibinfo {year} {2018})},\ \Eprint {http://arxiv.org/abs/1712.00688} {arXiv:1712.00688 [gr-qc]} \BibitemShut {NoStop}%
\bibitem [{\citenamefont {Lawrence}\ \emph {et~al.}(2023)\citenamefont {Lawrence}, \citenamefont {Turbang}, \citenamefont {Matas}, \citenamefont {Renzini}, \citenamefont {van Remortel},\ and\ \citenamefont {Romano}}]{Lawrence:2023buo}%
  \BibitemOpen
  \bibfield  {author} {\bibinfo {author} {\bibfnamefont {J.}~\bibnamefont {Lawrence}}, \bibinfo {author} {\bibfnamefont {K.}~\bibnamefont {Turbang}}, \bibinfo {author} {\bibfnamefont {A.}~\bibnamefont {Matas}}, \bibinfo {author} {\bibfnamefont {A.~I.}\ \bibnamefont {Renzini}}, \bibinfo {author} {\bibfnamefont {N.}~\bibnamefont {van Remortel}}, \ and\ \bibinfo {author} {\bibfnamefont {J.~D.}\ \bibnamefont {Romano}},\ }\href {\doibase 10.1103/PhysRevD.107.103026} {\bibfield  {journal} {\bibinfo  {journal} {Phys. Rev. D}\ }\textbf {\bibinfo {volume} {107}},\ \bibinfo {pages} {103026} (\bibinfo {year} {2023})},\ \Eprint {http://arxiv.org/abs/2301.07675} {arXiv:2301.07675 [gr-qc]} \BibitemShut {NoStop}%
\bibitem [{\citenamefont {Sah}\ and\ \citenamefont {Mukherjee}(2023)}]{Sah:2023bgr}%
  \BibitemOpen
  \bibfield  {author} {\bibinfo {author} {\bibfnamefont {M.~R.}\ \bibnamefont {Sah}}\ and\ \bibinfo {author} {\bibfnamefont {S.}~\bibnamefont {Mukherjee}},\ }\href@noop {} {\  (\bibinfo {year} {2023})},\ \Eprint {http://arxiv.org/abs/2307.06405} {arXiv:2307.06405 [gr-qc]} \BibitemShut {NoStop}%
\bibitem [{\citenamefont {Dey}\ \emph {et~al.}(2023)\citenamefont {Dey}, \citenamefont {Longo~Micchi}, \citenamefont {Mukherjee},\ and\ \citenamefont {Afshordi}}]{Dey:2023oui}%
  \BibitemOpen
  \bibfield  {author} {\bibinfo {author} {\bibfnamefont {R.}~\bibnamefont {Dey}}, \bibinfo {author} {\bibfnamefont {L.~F.}\ \bibnamefont {Longo~Micchi}}, \bibinfo {author} {\bibfnamefont {S.}~\bibnamefont {Mukherjee}}, \ and\ \bibinfo {author} {\bibfnamefont {N.}~\bibnamefont {Afshordi}},\ }\href@noop {} {\  (\bibinfo {year} {2023})},\ \Eprint {http://arxiv.org/abs/2305.03090} {arXiv:2305.03090 [gr-qc]} \BibitemShut {NoStop}%
\end{thebibliography}%

\end{document}